\documentclass[usenatbib]{mnras}

\usepackage{amssymb}
\usepackage{amsmath}
\usepackage{graphicx}
\usepackage{mathtools}
\usepackage{cuted}
\usepackage{epstopdf}

\newcommand{\eg}{E_{\rm g}}

\newcommand{\nbineg}{N_{\rm eg}}

\defcitealias{murga16a}{Paper~I}

\title[{\tt Shiva}: the dust destruction model]
{{\tt Shiva}: the dust destruction model}
\author[M. S. Murga et al.]
{M. S. Murga$^{1}$\thanks{E-mail: murga@inasan.ru},
 D. S. Wiebe$^{1}$, E. E. Sivkova$^{1}$, V. V. Akimkin$^{1}$\\
$^{1}$Institute of Astronomy, Russian Academy of Sciences, Pyatnitskaya str. 48, Moscow 119017, Russia}
\date{Accepted today. Received tomorrow; in original form \today}
\pubyear{2018}

\begin{document}
\label{firstpage}
\pagerange{\pageref{firstpage}--\pageref{lastpage}}\maketitle

\begin{abstract}
We present a numerical tool {\tt Shiva} designed to simulate the dust destruction in warm neutral, warm ionized, and hot ionized media under the influence of photo-processing, sputtering, and shattering. The tool is designed primarily to study the evolution of hydrogenated amorphous carbons (HACs), but options to simulate polycyclic aromatic hydrocarbons (PAHs), silicate and graphite grains are also implemented. HAC grain photo-processing includes both dehydrogenation and carbon atom loss. Dehydrogenation leads to material transformation from aliphatic to aromatic structure. Simultaneously, some other physical properties (band gap energy, optical properties, etc.) of the material change as well. The {\tt Shiva} tool allows calculating the time-dependent evolution of the dust size distribution depending on hydrogen, helium, and carbon number densities and ionization state, gas temperature, radiation flux, relative gas-dust and grain-grain velocities. For HAC grains the evolution of band gap energy distribution is also computed. We describe a dust evolution model, on which the tool relies, and present evolutionary time-scales for dust grains of different sizes depending on external conditions. This allows a user to estimate quickly a lifetime of a specific dust grain under relevant conditions. As an example of the tool usage, we demonstrate how grain properties and corresponding infrared spectra evolve in photo-dissociation regions, H{\sc II} regions, and supernova remnant shocks.
\end{abstract}


\section{Introduction}
A large number of infrared (IR) observations available to the date indicate that dust evolves in the interstellar medium (ISM). This evolution is related to both growth and destruction. Many studies have already been performed to show how these processes manifest themselves in various astrophysical environments. Dust grains are mostly destroyed in extreme conditions with enhanced ultraviolet (UV) radiation field (RF), shock waves, hot gas, or turbulent motions. At these conditions grains can be eroded by high energy photons \citep{jochims94, allain96_1,allain96_2}, via sputtering by fast ions and electrons \citep{draine79, tielens94}, or by shattering in grain-grain collisions \citep{jones96, hirashita09}. Rates of these processes depend on the dust material. While it is generally assumed that interstellar dust grains are most likely silicate and carbonaceous \citep{MRN, desert90, LD01, zubko04, jones13}, the specific nature of carbonaceous material in cosmic dust is still debated. In some models, which successfully explain observed dust extinction/IR emission properties, it is assumed that big carbonaceous grains are graphite-like and small carbonaceous grains are PAH-like \citep{DL07, siebenmorgen14}. In other models, it is assumed that carbon dust grains are hydrogenated amorphous carbons \citep{dustem, jones13}, with PAHs being their special case.

While non-evolving dust models are still successfully used to interpret available observations, various destructive processes for specific interstellar conditions and dust materials are also considered on a case-by-case basis. Given an ever increasing volume of observational data, it would be convenient to have a tool for a quick theoretical estimation of evolutionary time-scales and an analysis of a varying dust size distribution. With this in mind, we present a tool {\tt Shiva} that combines primary destructive processes into a single model and allows calculating the dust evolution under various conditions. The {\tt Shiva} tool can be used as a stand-alone feature or be included as a module into a more detailed hydrodynamical model, like the MARION model of an expanding H{\sc ii} region presented in the works of \citet{2015MNRAS.449..440A, 2017MNRAS.469..630A}. {\tt Shiva} shares many elements with the recently published THEMIS model~\citep{jones_themis}, however, in addition to processes presented in THEMIS (photo-aromatisation, sputtering), the {\tt Shiva} model also accounts for photo-destruction via carbon atom loss, aromatisation by ions and electrons, and shattering processes. Our numerical tool is publicly available\footnote{{\tt\url{http://www.inasan.rssi.ru/~khramtsova/SHIVA.html}}}.

The paper is organised as follows. In Sect.~\ref{sect: the model} we describe the main elements of the dust evolution model and present equations being solved. In Sect.~\ref{sect: timescales} we demonstrate some results obtained with the {\tt Shiva} tool. Finally, conclusions are given in Sect.~\ref{sect: summary}. In Appendix \ref{app: sput_coef} and \ref{app: photo_coef} we describe rate coefficients for photo-destruction and sputtering processes in more detail.

\section{The model}\label{sect: the model}

A model of cosmic dust evolution implemented in {\tt Shiva} includes photo-processing, i.e. hydrogen atom loss (aromatisation) and carbon atom loss, sputtering by ions and electrons, and shattering in grain-grain collisions. The model can be applied in a wide range of astrophysical objects except for those, where accretion/coagulation processes play an important role (i.e. molecular clouds, protoplanetary discs). The main concepts of the model were described in the works of \citet[][hereafter Paper I]{murga16a} and \citet{murga16b}. In this work we present an advanced version of the model which now includes aromatisation by ions and treatment of grain charge, employs a dedicated method for solution of stiff kinetic equations, and is publicly available. Below we recapitulate the features of the model, paying more attention to its modifications. Processes of sputtering and shattering are treated analogously to Paper~I, so details can be found there. For reference, expressions for sputtering rates are given in Appendix \ref{app: sput_coef}.

The key element of the model is consideration of aromatisation process of carbonaceous grains, that is, restructuring of dust grains from the aliphatic state to the aromatic state as a result of dehydrogenation. Alteration of the material structure leads to a change of its band gap energy, $E_{\rm g}$, which decreases as a grain becomes more aromatic. Other considered processes change grain mass. Thus, each grain can be described by two parameters, its mass $m$ (or an equivalent radius $a$, which we calculate assuming a compact grain with solid density of 1.5~g~cm$^{-3}$ as appropriate for HAC material) and the band gap energy $\eg$. These parameters evolve under the influence of external factors. We divide the mass interval [$m_{\rm{min}}$, $m_{\rm max}$] into $N_{\rm m}$ bins with $N_{\rm m}+1$ borders, $m_{\rm b}^{i}$. The band gap interval [$\eg^{\rm min}$, $\eg^{\rm max}$] is divided into $\nbineg$  bins with $\nbineg+1$ borders, $E_{\rm gb}^{j}$. Thus, the total number of bins is $N_{\rm m} \times \nbineg$. Each bin can be characterised by the grain number density $N_{ij}$, cm$^{-3}$, which is the number of dust grains within the bin, so that the total dust number density is $\sum_{i=1}^{N_{\rm m}}\sum_{j=1}^{N_{\rm eg}}N_{ij}$. We assume that all dust grains are initially in the same aromatisation state. If the initial mass distribution $dn/dm$ is known, $N_{ij}$ can be expressed as $dn/dm(m_{\rm b}^{i+1}-m_{\rm b}^{i})$. The evolution of $N_{ij}$ can be written in a form of the discrete Smoluchowski equation \citep{smolukh} that is often used for the description of coagulation process \citep{dullemond05,akimkin15a}:

\newpage
\begin{strip}
\begin{equation}
\begin{split}
\frac{dN_{ij}}{dt} &= \underbrace{A_{ij+1}^{(1)}N_{ij+1} - A_{ij}^{(1)}N_{ij}}_{\text{aromatisation by photons}} + \underbrace{B_{i+1j}^{(1)}N_{i+1j} - B_{ij}^{(1)}N_{ij}}_{\text{photo-destruction}}  \\
 &+ \underbrace{A^{(2)}_{ij+1}N_{ij+1} - A^{(2)}_{ij}N_{ij}}_{\text{aromatisation by ions in inelastic collisions}} + \underbrace{B^{(2)}_{i+1j}N_{i+1j} - B^{(2)}_{ij}N_{ij}}_{\text{sputtering due to inelastic collisions}}  \\
&+\underbrace{A^{(3)}_{ij+1}N_{ij+1} - A^{(3)}_{ij}N_{ij}}_{\text{aromatisation by ions in elastic collisions}}
+\underbrace{B^{(3)}_{i+1j}N_{i+1j} - B^{(3)}_{ij}N_{ij}}_{\text{sputtering due to elastic collisions}} \\
&+ \underbrace{A^{(4)}_{ij+1}N_{ij+1} - A^{(4)}_{ij}N_{ij}}_{\text{aromatisation of big grains due to sputtering}}+ \underbrace{B^{(4)}_{i+1j}N_{i+1j} - B^{(4)}_{ij}N_{ij}}_{\text{sputtering of big grains}}\\
&+\underbrace{\frac{1}{2}\sum_{i_1=1}^{N_{\rm m}}\sum_{j_1=1}^{\nbineg}\sum_{i_2=1}^{N_{\rm m}}\sum_{j_2=1}^{\nbineg} C_{i_1j_1i_2j_2}^{ij} L_{i_1i_2} N_{i_1j_1}N_{i_2j_2} - N_{ij}\sum_{i_1=1}^{N_{\rm m}}\sum_{j_1=1}^{\nbineg}L_{i_1i} N_{i_1j_1}}_{\text{shattering}}  \\ 
\end{split}
\label{eg: big formula}
\end{equation}
\end{strip}

\noindent Here $A_{ij+1}^{(1,2,3,4)}$ are rate coefficients that describe a hydrogen atom loss and a corresponding increase in aromatisation in the direction from a $(j+1)$th bin to a neighbouring $j$th bin by $\eg$ (or from $j$th to $(j-1)$th bin in case of $A_{ij}^{(1,2,3,4)}$). We assume that only a loss of carbon atoms leads to the change in grain mass, so that the mass bin of a grain is not changed due to aromatisation process. $B_{i+1j}^{(1,2,3,4)}$ are rate coefficients that describe the carbon atom loss by photo-destruction and sputtering and a corresponding change in grain mass from an $(i+1)$th bin to a neighbouring $i$th bin (or from $i$th to $(i-1)$th bin in case of $B_{ij}^{(1,2,3,4)}$). We assume that the loss of carbon atoms does not change the grain aromatisation state. $C_{i_1j_1i_2j_2}$ is a ratio between the mass transferred to $i$th bin due to shattering and the average mass of the bin (see other details in Paper I), and $L_{{i_1}{i_2}}$ is the shattering kernel:
\begin{equation}
L_{{i_1}{i_2}} = \pi(a_{i_1}+a_{i_2})^2 v^{\rm col}_{{i_1}{i_2}} F_{\rm C}
\end{equation}
with $v^{\rm col}_{{i_1}{i_2}}$ being the velocity of a grain-grain collision and $F_{\rm C}$ being the Coulomb factor, which is estimated as
\begin{equation}
F_{\rm C} = 1 - \frac{2 Z_{i_1}Z_{i_2} e^2}{m_{i_1i_2}(a_{i_1}+a_{i_2}) (v^{\rm col}_{{i_1}{i_2}})^2},
\end{equation}
where $Z_{i_1}$ and $Z_{i_2}$ are charge numbers of grains, $e$ is electron charge, $m_{i_1i_2}$ is a reduced mass of colliding grains. The value of $F_{\rm C}$ is set to zero if the above expression gives negative values.

The rate coefficients of aromatisation are calculated as
\begin{equation}
A^{(1,2,3,4)}_{ij} = \frac{\varepsilon^{(1,2,3,4)}_{ij}}{E_{{\rm gb}}^{j+1}-E_{{\rm gb}}^{j}}.
\end{equation}
Boundary values of $A^{(1,2,3,4)}_{ij}$ for $j=1$ and $j=N_{\rm eg}+1$ are set to zero. 

The rate coefficients for carbon atom loss are calculated analogously:
\begin{equation}
B^{(1,2,3,4)}_{ij} = \frac{\mu^{(1,2,3,4)}_{ij}}{m_{\rm b}^{i+1}-m_{\rm b}^{i}}.
\end{equation}
Boundary values of $B^{(1,2,3,4)}_{ij}$ for $i = N_{\rm m}$ are set to zero, but we allow grains to leave the bin with $i=1$. We define $\varepsilon^{(1,2,3,4)}_{ij}$ and $\mu^{(1,2,3,4)}_{ij}$ in the following sections. 

The system of ordinary differential equations (ODE) (\ref{eg: big formula}) is stiff and should be solved using an appropriate method. We chose the solver {\tt CVODE}\footnote{{\tt\url{https://computation.llnl.gov/projects/sundials/cvode}}} from the {\tt SUNDIALS} software package~\citep{cvode} that allows solving stiff systems of ODE. 

The {\tt Shiva} code is implemented as a C$^{++}$ computer code. Detailed instructions for users can be found in the web link specified above.  

\subsection{Photo-processing}\label{sect: photodestruction}

We consider photo-destruction processes that include the loss of both carbon and hydrogen atoms. In general, the absorption of a UV photon by a grain can lead to several outcomes: ionization~(charging), IR photon emission, dissociation of C--H or C--C bonds. These processes occur with probabilities $Y_{\rm ion}$, $Y_{\rm IR}$, $Y_{\rm diss}^{\rm CH}$, $Y_{\rm diss}^{\rm CC}$, respectively, which sum up to unity. If we assume that photon energy is spent to dissociation, then a C--H bond is more likely to be destroyed as it is less strong than a C--C bond. Thus, a dust grain first loses its hydrogen atoms and becomes partially or fully dehydrogenated, so that $Y_{\rm diss}^{\rm CC}=0$ at this stage. Simultaneously the structure of a grain material changes from aliphatic to aromatic\footnote{This applies only to HAC material.} as is described in detail in the works of \citet{jones12_1, jones12_2, jones12_3}. In these papers it is mentioned that the maximum band gap energy value is 2.67~eV, and the minimum value depends on a grain radius, but we set this value equal to 0.1~eV for all dust grains as it was adopted for aromatic material in the work of \citet{jones13}. The rate of change of the band gap energy $\varepsilon^{(1)}_{ij}$ due to aromatisation by photons can be found in Appendix \ref{app: photo_coef}.

When a grain is dehydrogenated down to $\eg=0.1$~eV (or {is a PAH particle}), another destruction scheme with carbon atoms loss starts to operate, and at this stage we assume $Y_{\rm diss}^{\rm CH}=0$. This process is relevant only for small particles, so we limit its action to grains with the number of carbon atoms less than 1000. The loss of carbon atoms requires more energy and apparently is only possible near a strong UV radiation source \citep{jochims94, allain96_1, zhen15}. In the model we assume that subsequent dissociation proceeds analogously to PAHs as described in the work of \citet{allain96_1}\footnote{It should be mentioned that small HAC particles described by \citet{jones13} evolve with time, and are very likely aromatic in the ISM. However, even small aromatic grains are not identical to PAHs, as they are volumetric in a general case, though in some specific cases they can be PAHs \citep{jones90}. But we assume that dissociation properties (like the binding energy) for small grains are always  the same as for PAHs.}. Expression for dissociation probability of dehydrogenated grains $Y_{\rm diss}^{\rm CC, sp}$ can be found in Appendix \ref{app: photo_coef}.

In the model we consider the multi-photon heating mechanism described in the work of \citet{guhathakurta89}. It implies that a dust grain can accumulate energies of several absorbed photons before this energy is spent on the grain IR emission. The internal energy of a grain can grow several times higher in this case than in a single-photon approach, and, thus, dissociation occurs more effectively. We calculate the energy history for a grain with a radius $a$ immersed in the radiation field with the flux $F(E)$ and then derive the probability distribution of its internal energy $p(E)$ using the approach described in the work of \citet{pavyar12}. Importance of the multi-photon approach for small grain destruction has been noted in the work of \citet{visser07}. Thus, the dissociation probability $Y_{\rm diss}^{\rm CC, sp}$ should be integrated over all possible internal energies:
\begin{equation}
Y_{\rm diss}^{\rm CC}(a_i, Z_i) = \int\limits_{0}^{\infty}Y_{\rm diss}^{\rm CC, sp}(a_i,E, Z_i)p(E)dE.
\end{equation}  

The rate of a grain mass decrease as a result of photo-destruction is expressed as
\begin{equation}
\mu_{ij}^{(1)} = \frac{\mu_{\rm d}}{N_{\rm A}} Y_{\rm diss}^{\rm CC}(a_i, Z_i)\int\limits_0^{\infty}  Q_{\rm abs}(a_i,E, Z_i) a_i^2 \frac{F(E)}{E} dE,
\end{equation} 
where $Q_{\rm abs}$ is a grain absorption efficiency coefficient, $F(E)$ is the incident radiation flux, erg~cm$^{-2}$~s$^{-1}$~eV$^{-1}$, $E$ is photon energy, $\mu_{\rm d}$, g~mol$^{-1}$, is molar weight of the grain material, $\mu_{\rm d} = 12-11 X_{\rm H}$, $X_{\rm H}$ is hydrogen atom fraction, and $N_{\rm A}$ is Avogadro constant. The ratio of $\mu_{\rm d}$ and $N_{\rm A}$ is used to convert the rate from the number of atoms per second to gram per second. In the current study we assume that the radiation field is represented by the scaled interstellar radiation field (ISRF), taken from \citet{MMP83}, which we refer to as MMP83. The scale factor is $U$, so that the local radiation flux is $U\times{\rm MMP83}$. Other spectra $F(E)$ can be used as well. The adopted optical properties (in particular, $Q_{\rm abs}$) are discussed in subsection \ref{sect: charge role}.

Photo-destruction of carbonaceous grains is in fact a blurred question. Partly this is related to the lack of knowledge of which types of grains are encountered in space. Since the unidentified infrared (UIR) bands had been discovered \citep[e.g.][]{1973ApJ...183...87G}, further investigations have constrained possible assumptions, but the final clarification is yet to come. While it is more or less reliably established that small carbonaceous grains should contain at least some aromatic rings to explain the UIR bands~\citep{allamandola85}, their actual structure, size, chemical formulae are debated. Several candidates that are considered to date are PAHs~\citep{leger84, allamandola85}, HAC grains \citep{duley85,jones13}, mixed aromatic-aliphatic organic nanoparticles~\citep{kwok11}. Even if we adopt the most studied candidates, PAHs, as typical representatives of the smallest grains, their exact shape is also far from being fully understood. While some studies indicate that PAHs should be compact, pericondensed and consist of no less than 30--40 atoms \citep{allamandola89, jochims94, langhoff96}, other works offer different conclusions~\citep{ekern98}, stressing that the photo-destruction rate strongly depends on many parameters and not only on grain size and shape. It has been shown that biphelynene (C$_{12}$H$_{8}$) is a stable molecule in spite of consisting of only two aromatic rings. Coronene (C$_{24}$H$_{12}$) loses hydrogen atoms under the influence of UV radiation, but catacondensed chrysene or tetracene (both C$_{18}$H$_{12}$) lose one or two hydrogen atoms and acetylenic group and then become photostable. So, apparently, there is no universal approach to the calculation of photo-destruction rates and products for any PAHs, and it is necessary to consider each molecule individually. 

The situation is not much clearer with the photo-destruction of HAC grains. Many experiments prove that their dehydrogenation occurs under UV irradiation \citep{mennella01}. Rates and products of HAC photo-destruction were studied in recent laboratory experiments \citep{alata14, alata15}, but perhaps these rates cannot be applied directly to the ISM. Interstellar HAC grains are probably more aromatic, at least on the surface, and consequently are more stable. So in the absence of a general theory for the photo-destruction of HAC grains, at the first destruction step we apply experimental data, if the grains are predominantly aliphatic, and the RRK (Rice-Ramsperger-Kassel) theory, adopted to PAHs by  \citet{allain96_1}, if the grains are aromatic.

\subsection{Grain charge} \label{sect: charge role}

Supplementing our previous work, we include a charge of a grain $Z$ in the calculations of the destruction rates. It was shown that charged and neutral small grains have different absorption properties~\citep{1977CaJPh..55.1930B, 2012OptL...37..265K,  2012JQSRT.113.2561K}. Moreover, the fate of an absorbed photon~(ionization, IR photon emission, dissociation of C--H and C--C bonds) also depends on a grain charge. In this paper we study how the charge influences branching over these possible channels. We consider only variations in $Q_{\rm abs}$ in the mid-IR range and not in the UV range, so the main factor is a grain cooling rate, not a grain heating rate.

The charge state of dust grains may affect the efficiency of the processes considered in this work. Quite naturally, even large charge numbers are not important for shattering as at high collision velocities needed for the fragmentation the Coulomb factor is close to unity. Sputtering of big grains is also insensitive to their charge. However, charged and neutral small grains do respond differently to photo-destruction and sputtering. First, small ionized grains have different optical properties that affect their IR-emission rate. Under non-equilibrium conditions, when single-photon heating is important, IR opacities define the rate, with which the grain sheds the extra energy, and destruction probability. Second, it is more difficult to ionize grains that are already positively charged. This is why the dissociation yields are different for ionized and neutral grains. Here we consider how the grain charge influences the optical properties. Comparison of various rates for ionized and neutral grains is presented in Sect.~\ref{subsect: charge rates}.

We adopt optical properties for neutral dust grains presented in the works of \citet{jones12_1, jones12_2, jones12_3} with some modifications concerning the bands around $7-8\,\mu$m described in the work of \citet{jones13}. The full list of mid-infrared bands, corresponding to both aliphatic and aromatic bonds, can be found in Table~C.1 of the work of \citet{jones13}. Modified properties for dehydrogenated HAC were calibrated to the properties used in the DustEm model \citep{dustem}.

To compare efficiencies of destructive processes for charged and neutral grains one needs to know optical properties for charged grains as well. Unfortunately, studies of optical properties of ionized HAC grains are apparently lacking, while the properties of PAH cations are well investigated. In this situation the only way to proceed is to assume that ionization alters optical properties of HAC grains in the same way as it changes optical properties of PAHs. Perhaps, this approach is justified as most small dust grains in the ISM should be aromatised \citep{jones13}, and PAHs can be considered as a special case of small aromatic HACs. Moreover, we only need to know the properties of charged grains, when we calculate rates of those processes in which the smallest grains are involved. Thus, we describe the properties of such ionized grains as if they were PAHs, while the optical properties of bigger grains are assumed to be independent on the charge state. In this work we deal with conditions where the dust grains are either neutral or positively charged, so we consider only cations. Also, we do not take into account multiple ionized states in calculations of optical properties because in this case we would have to make too many assumptions. So, talking about optical properties of charged dust grains, we only mean singly ionized cations.

It has been shown \citep{pauzat93, schutte93, langhoff96, boersma11} that charge has a strong influence on aromatic infrared bands. Specifically, intensities of C--H bands (around $3.3\,\mu$m) are reduced \citep{langhoff96}, while intensities of C--C stretching and in-plane C--H bending modes ($6-9\,\mu$m) are enhanced for ionized PAHs \citep{szczepanski93, langhoff96, hudgins99, kim01, bauschlicher09}. The bands corresponding to C--H out-of-plane bending are also shifted due to the charge \citep{hudgins99, boersma13, shannon16}. Based on these and other works, the optical properties for both neutral and ionized PAHs were reproduced in the PAH model presented in the works of \citet{LD01, DL07}. We adopt the band variations related to the ionized state analogously to these works. In particular, we make the following modifications:
\begin{itemize}
\item The integrated strength of C--H stretching bands (3.25, 3.28, 3.31, 3.32, 3.35, 3.38, 3.42, 3.45, 3.47, 3.48, $3.51\,\mu$m) for ionized grains is smaller by a factor of 4.4 compared to neutral grains.
\item The integrated strength of aromatic C--C stretching bands (6.10, 6.25, 6.67, 7.53, 7.69, 7.85, $8.60\,\mu$m) for ionized grains is larger by a factor of 8 compared to neutral grains.
\end{itemize} 
Other modes are left unchanged.

\begin{figure*}
	\includegraphics[width=0.45\textwidth]{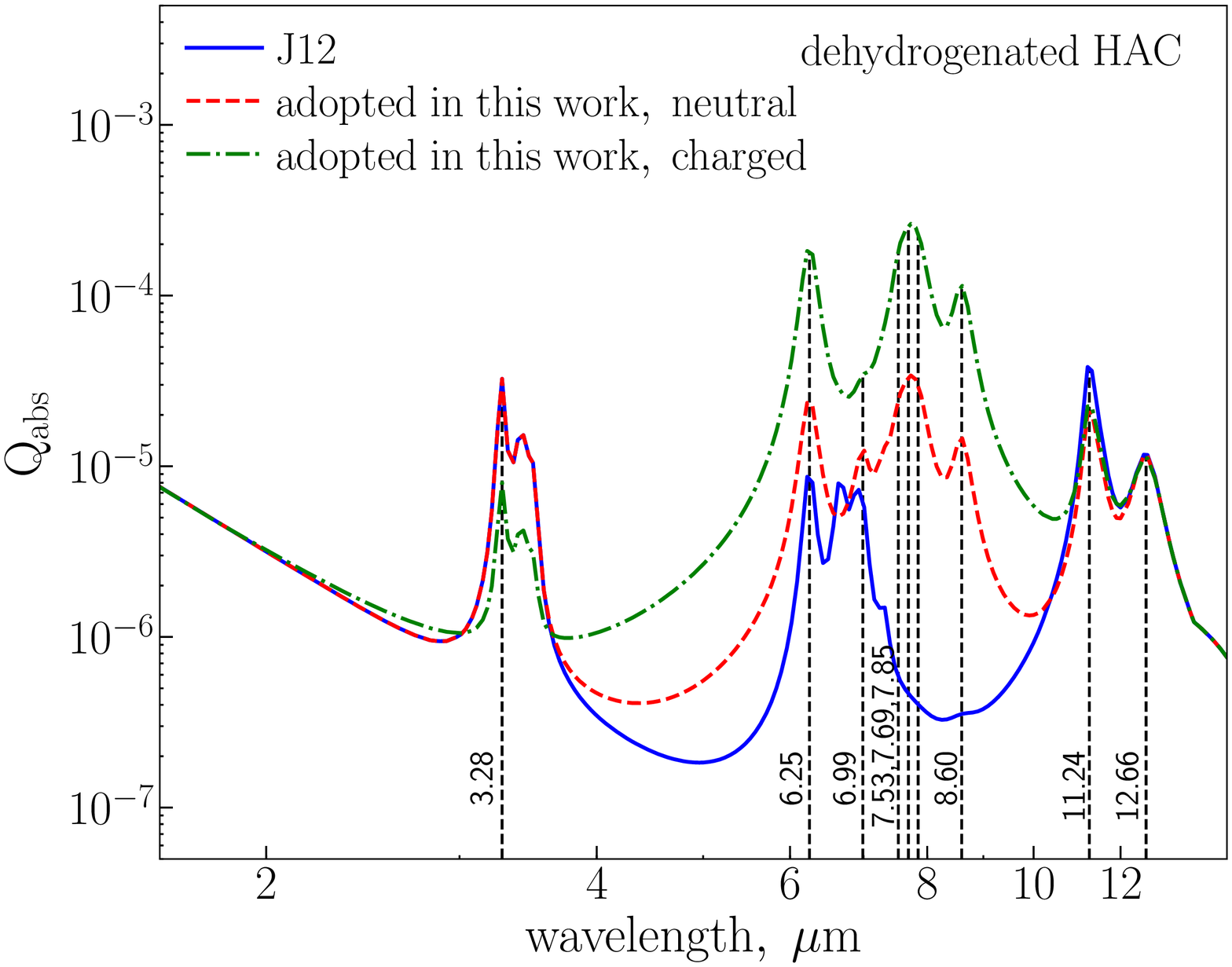}
	\includegraphics[width=0.45\textwidth]{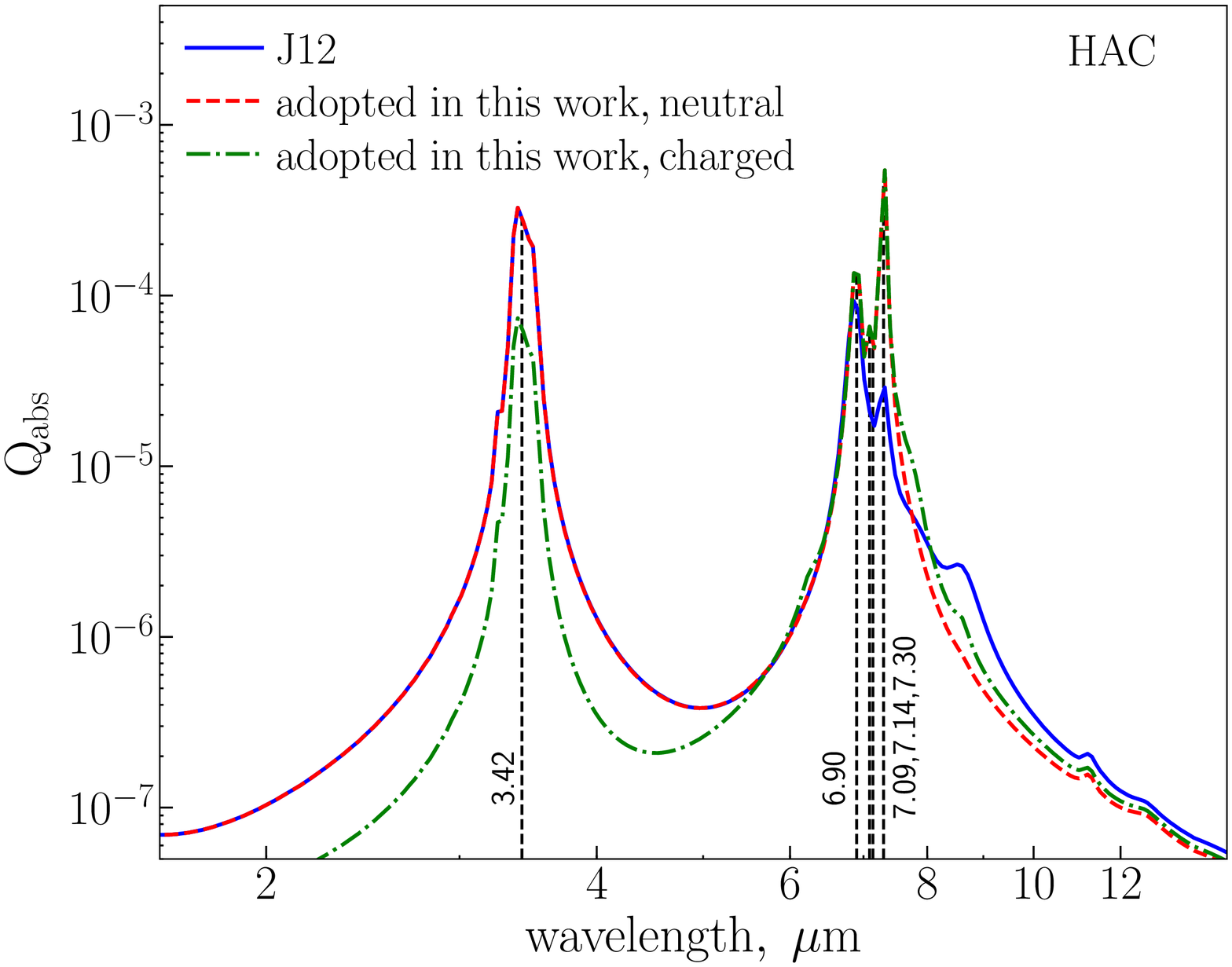}
	\caption{Optical properties of HAC grains with a size of 5~\AA{} and $\eg=0.1$~eV (left panel) and $\eg=2.67$~eV (right panel). The original neutral grain properties from the work of \protect{\citet{jones12_3}} are shown with blue lines (labelled as J12), the adopted properties for neutral grains modified according to the work of \protect{\citet{jones13}} are shown with green lines, and the adopted properties for charged grains are shown with red lines. The main features corresponding to aromatic and aliphatic modes are marked on each panel with dashed lines.}
	\label{qabs}
\end{figure*}

In Fig.~\ref{qabs} we demonstrate how the adopted optical properties differ from the original properties from the work of \citet{jones12_3} for the grain size of 5~\AA{} and the band gap energies 0.1~eV (left panel) and 2.67~eV (right panel). These $\eg$ values correspond to aromatic (dehydrogenated) and aliphatic (hydrogenated) states, respectively. We focus on the wavelength range from 2 to 15~$\mu$m assuming that properties are identical in other spectral bands. The original data from \citet{jones12_3} are shown with blue lines. Green lines correspond to a newer version of optical data from \citet{jones13}, where some bands around $8\,\mu$m have been added. We utilise these data for neutral grains. The adopted $Q_{\rm abs}$ for charged grains are shown with red lines.

In the presented version of the model a grain charge is treated as an input parameter for each bin and should be specified from some separate considerations.

\section{Results}\label{sect: timescales}

Here we present some results produced by the {\tt Shiva} tool. We show how the grain charge influences grain destruction rates in subsection \ref{subsect: charge rates}. Then we apply the model to some specific astrophysical conditions. In the general ISM both the ISRF and relative motions of gas and dust only slightly affect the distributions of sizes and aromatisation states of dust grains. But there are more extreme objects, like photo-dissociation regions~(PDRs), H{\sc ii} regions, and supernova remnants, where dust is exposed to more powerful destructive  factors. For clarity, we consider separately the extreme medium with the strong radiation field and the extreme medium with high velocities and temperatures. We follow the evolution of dust grains in these two environments in subsections \ref{subsect: PDR} and \ref{subsect: SNR}.

\subsection{Rate comparison for charged and neutral grains}\label{subsect: charge rates}

It was mentioned above that charge plays a role in defining the cooling rate of a grain and its ionization probability. Both these quantities are needed to estimate the photo-destruction rate. Here we consider how they differ for neutral and charged grains and how this difference influences the photo-destruction rate.

\begin{figure*}
	\includegraphics[width=0.45\textwidth]{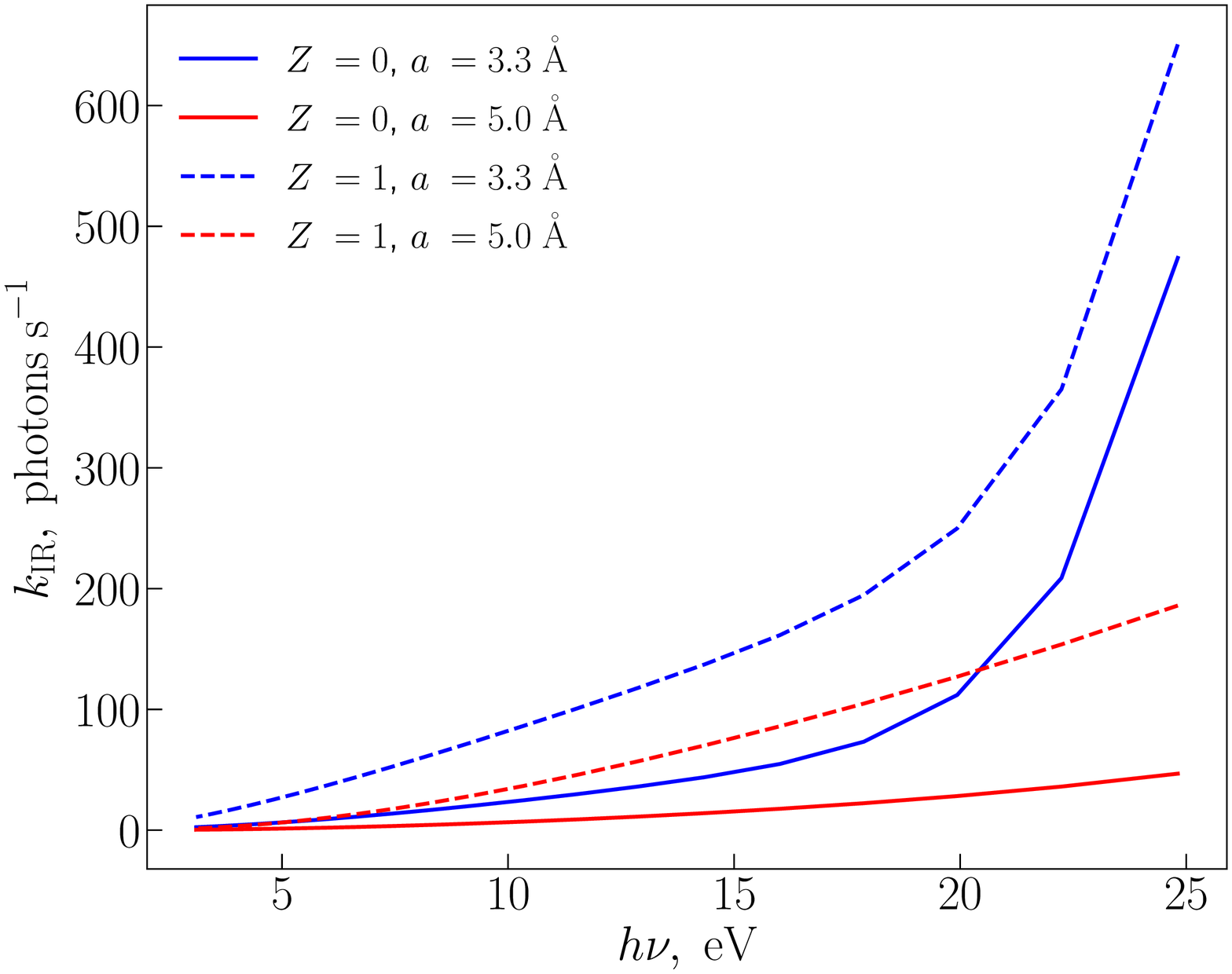}
	\includegraphics[width=0.45\textwidth]{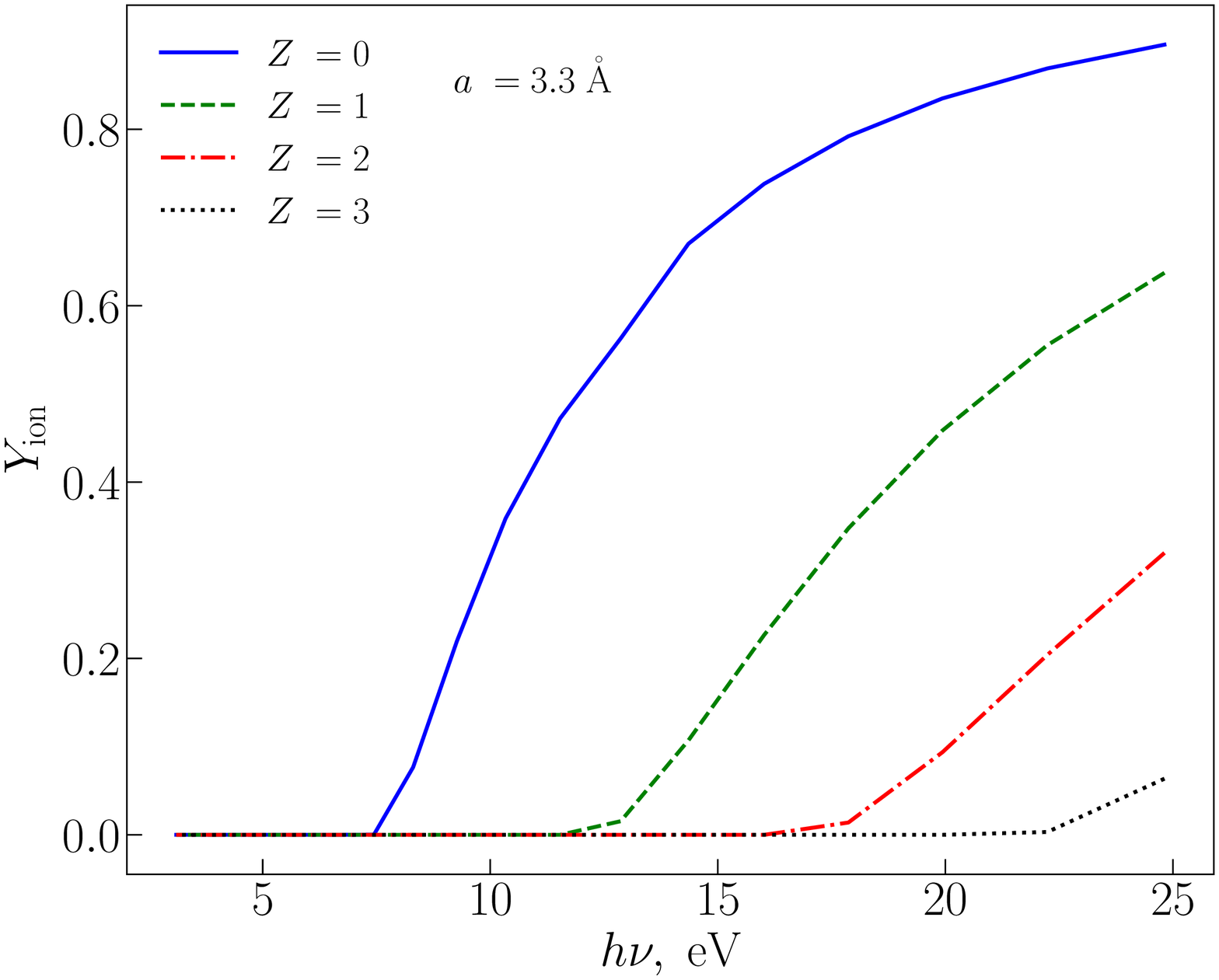}
	\caption{On the left: dependence of the IR emission rate on the photon energy. Blue solid and dashed lines correspond to 3.3~\AA{} grains with zero and unit charges, respectively. Red solid and dashed lines show the same data for 5~\AA{} grains. On the right: dependence of the ionization yield on the photon energy for a 3.3~\AA{} grain with charge numbers of 0~(blue), 1~(green), 2~(red), 3~(black).}
	\label{kir_Yion}
\end{figure*}

Fig.~\ref{kir_Yion}~(left) illustrates the dependence of $k_{\rm IR}$ on the photon energy for ionized and neutral grains with radii of 3.3 and 5~\AA{}. We further assume that absorption efficiency $Q_{\rm abs}$ for charged HAC grains  is determined by the very presence of non-zero charge, but does not depend on its value or sign. So we compare the IR emission rates ($k_{\rm IR}$) only for HAC$^0$ and HAC$^{+}$. For both sizes $k_{\rm IR}$ is several times higher for ionized grains, and this difference increases with the photon energy. This implies that a neutral grain is less stable to photo-destruction than a charged grain, because its IR relaxation rate is lower.

In Fig.~\ref{kir_Yion}~(right) we show the dependence of the ionization yield $Y_{\rm ion}$ computed using an approach from the work of \citet{wd01_ion} on the photon energy for a 3.3~\AA\ grain with charge numbers of $Z = 0,1,2,3$. Obviously, $Y_{\rm ion}$ decreases with increasing charge. In other words, probability to ionize the grain drops with each subsequent ionization. Therefore, more energy is available for destruction or emission. In this respect, an ionized grain is less resistant to destruction than a neutral one, contrary to what is said  in the previous paragraph.

\begin{figure}
	\includegraphics[width=0.45\textwidth]{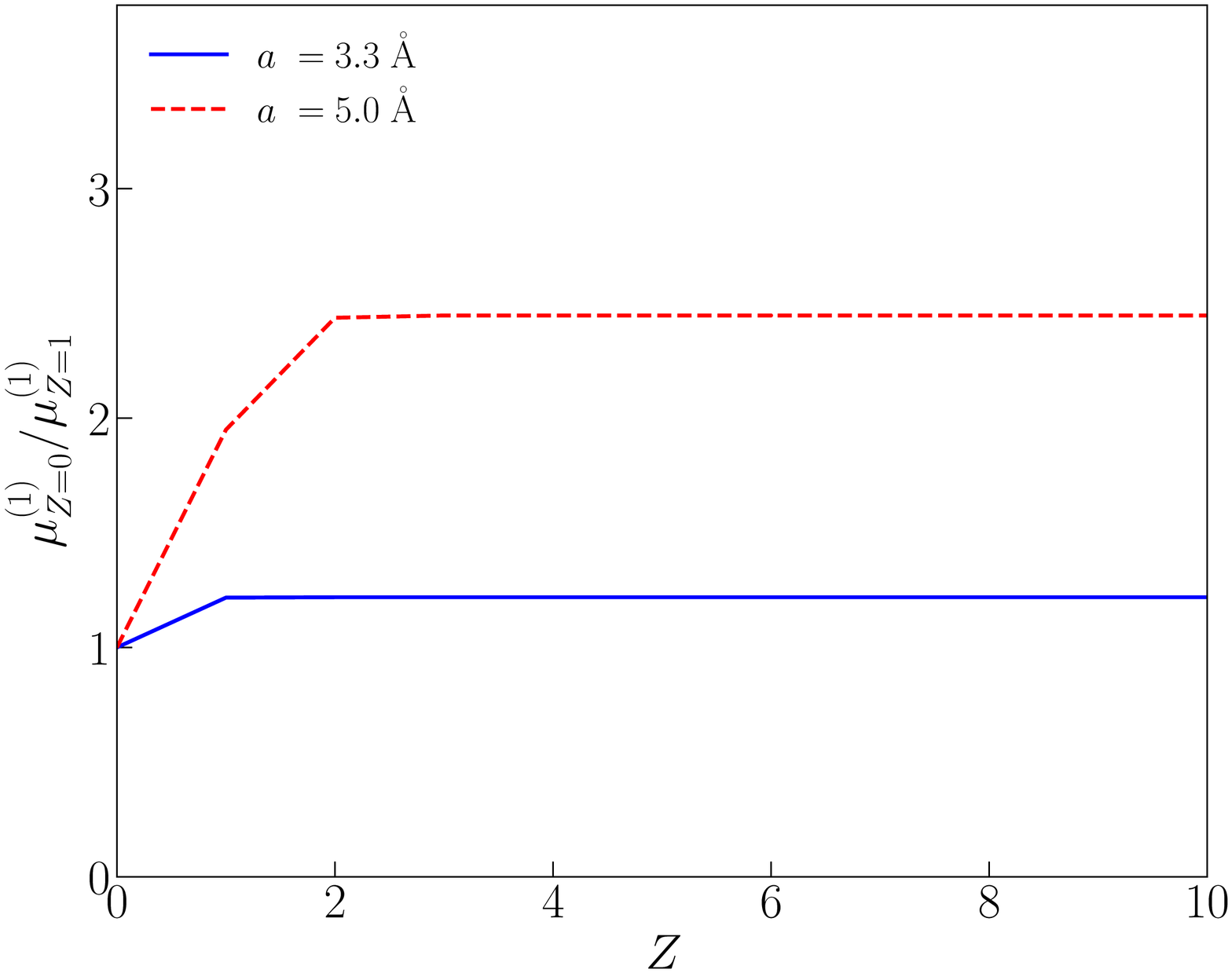}
	\caption{A ratio between photo-destruction rates of a neutral grain and a positively charged grain for various charge values. The blue line corresponds to a grain radius of 3.3~\AA{}, and the red line corresponds to a grain radius of 5~\AA.}
	\label{rate_Z}
\end{figure}

Relying upon calculations of photo-destruction rates for charged and neutral grains and taking into account the combined dependence of $Y_{\rm ion}$ and $k_{\rm IR}$ on the charge state, we conclude that the net effect makes charged grains less stable to photo-destruction than neutral ones, which is further demonstrated in Fig.~\ref{rate_Z}. For a 3.3~\AA{} grain the rate is increased by a factor of $\sim1.2$, while for the 5~\AA{} grains the factor can be as high as $\sim2.5$, depending on the charge value. These ratios are independent on the RF intensity. The larger are dust particles, the greater is the difference between the photo-destruction rates for ionized and neutral grains, but the rates themselves are too small to influence the relevant evolutionary time-scales. The analogous conclusions were made by \citet{allain96_2}, where they investigated the stability of PAHs. We obtain a similar difference for the sputtering rate due to elastic interaction with fast moving ions/electrons.

\subsection{Evolution of dust in the medium with enhanced UV radiation field}\label{subsect: PDR}
\subsubsection{Time-scales of photo-processing}

Before we proceed to further results, let us introduce two time-scales, which we consider as characteristic evolutionary times for various processes. We define a time-scale $\tau$ as a time needed to reach an $e$-fold change in the number density in an $i$th size bin summed over all aromatisation bins ($\sum\limits_{j}N_{ij}(t)$) relative to the initial density $\sum\limits_{j}N_{ij}(0)$. This time-scale is related to the overall evolution of the grain size distribution. Its value can be both positive and negative. If the number density in a bin decreases with time, $\tau$ is positive and represents the time needed to reduce the number density by a factor of $e$. Conversely, if the number density increases with time, then $\tau$ is negative. Another time-scale, $\tau_{\rm a}$, is the time during which the number density $N_{ij_{\rm min}}(t)$ in the $i$th mass bin and the $(j_{\min})$th aromatisation bin (the highest aromatisation state) becomes $\approx e$ times larger than the number density $N_{ij_{\rm max}}(t)$ in the same mass bin $i$, but in the aromatisation bin $j_{\rm max}$ (the lowest aromatisation state). In this study, we assume that initially all the aromatisation bins, except for bins with $\eg=\eg^{\max}$, are empty. This time-scale roughly characterises the aromatisation rate in a given mass bin.

\begin{figure}
	\includegraphics[width=0.45\textwidth]{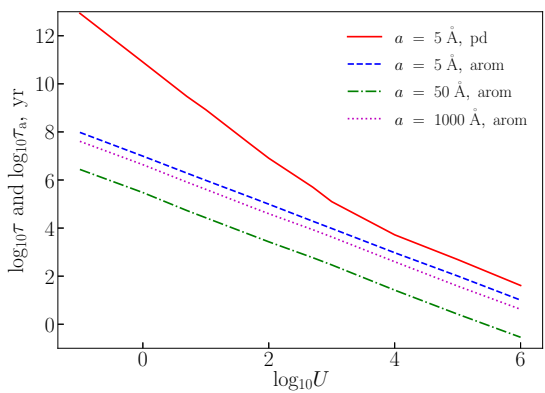}
	\caption{Dependence of the photo-destruction and aromatisation time-scales on the scaled radiation field, $U$. The photo-destruction time-scale of a 5~\AA{}-grain is shown with a red line. Aromatisation time-scales for 5, 50 and 1000~\AA{}-grains are shown with blue, green, and magenta lines, correspondingly.}
	\label{Utimescales}
\end{figure}

When the radiation prevails over other factors, we can consider only the photo-destruction processes (hydrogen and carbon loss). In Fig.~\ref{Utimescales} the dependence of the photo-destruction and aromatisation time-scales on the scaled radiation field, $U$, is shown. The red line shows the time-scale of a 5~\AA{}-grain destruction due to carbon loss under the influence of the UV field. We see that the destruction time-scale for such grains drops below $10^6$~yr only when $U$ exceeds $10^3$. For example, in the Orion Bar, where $U$ is estimated to be $10^4-10^5$~\citep{tielens85}, PAH-like grain should be destroyed on a time-scale of about $10^3$ yr, which seems to be  relevant for this PDR \citep{2006ApJ...637..823K}. In another PDR, Horsehead Nebula, the value of $U$ is smaller, about $10^2$~\citep{compiegne07}. In this case the PAH-like grains (or, at least, their carbon skeletons) should survive for a longer time, up to several Myr. The rate of hydrogen loss is greater, and the full dehydrogenation of aromatic species can lead to formation of fullerenes \citep{berne12}, but in this work the corresponding evolutionary path is not considered. Also, we do not consider dehydrogenation of PAHs and HAC grains with $\eg<0.1$~eV as was pointed above.

Blue, green, and magenta lines in Fig.~\ref{Utimescales} illustrate the aromatisation time-scales of 5, 50 and 1000~\AA{}-grains, correspondingly. It can be seen that the time-scale for each grain is inversely proportional to the intensity of the radiation field. Among the considered sizes, aromatisation is most effective for 50~\AA{}-grains. Analogous results were obtained in the work of \citet{jones14}, which we rely upon in our calculations. This result comes from Eq.~\ref{eq: arom}. Absorption efficiency grows with size while other values are constant until the size reaches 200~\AA{}, as small grains are in the Rayleigh regime, so the aromatisation rate increases with size. At $a=200$~\AA{}, $Y_{\rm diss}^{\rm CH}$ becomes inversely proportional to the grain size, while the absorption efficiency is independent on the size, because large grains are in the geometrical regime, so the time-scale of aromatisation for large grains (in our case 1000~\AA{}-grains) turns out to be smaller than that for small grains (5~\AA{}). 

\begin{figure*}
	\includegraphics[width=0.45\textwidth]{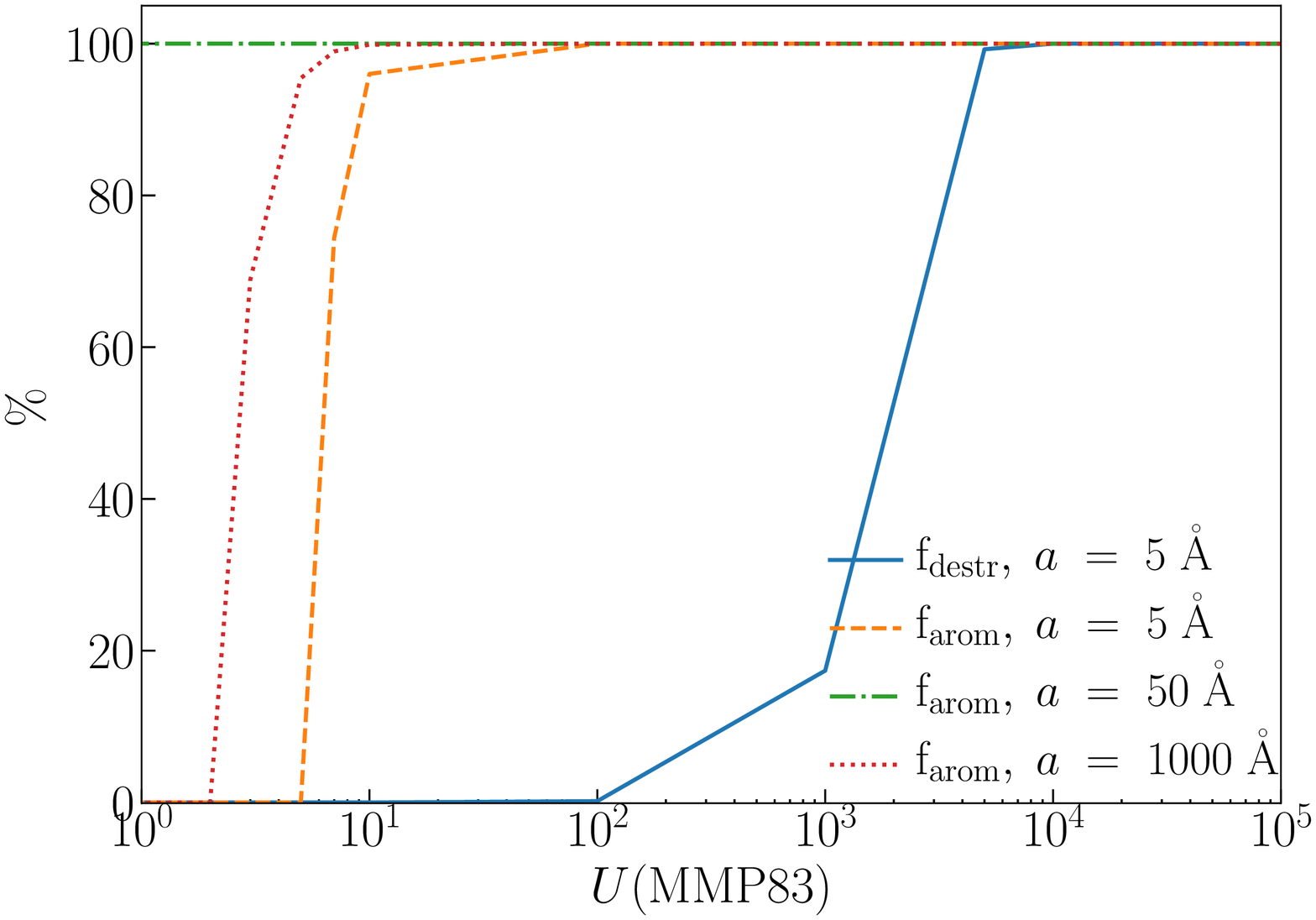}
	\includegraphics[width=0.45\textwidth]{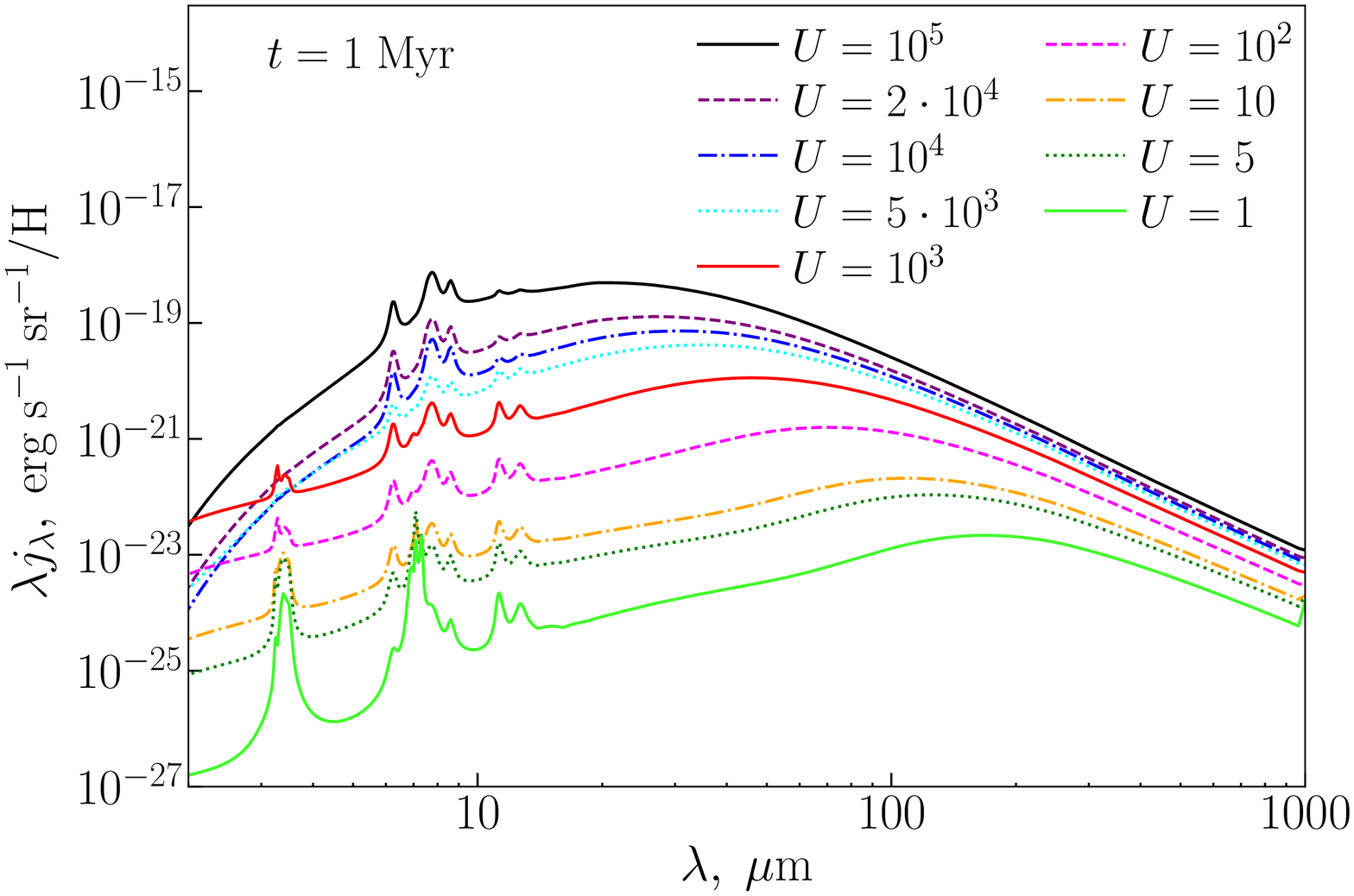}
	\caption{On the left: a fraction of photo-destroyed dust grains (blue line) and a fraction of aromatised dust grains with radii 5~\AA{}~(orange), 50~\AA{}~(green), 1000~\AA~(red) as functions of the radiation field intensity {after 1 Myr of evolution}. On the right: infrared spectra of dust exposed to the UV radiation with intensities in the range from 1 to $10^5$ of MMP83.}
	\label{frac_destr}
\end{figure*}

\subsubsection{Evolution of grains and infrared spectra in PDRs}

In Fig.~\ref{frac_destr} we present results of calculations that include only photo-processes for HAC particles with the initial J13 size distribution~\citep{jones13}, assuming that all grains are initially hydrogenated. We took the evolution time interval of 1~Myr which is a typical lifetime of PDRs. The considered intensities are varied from 1 to 10$^5$~MMP83. We take into account the charge of dust grains, which is calculated using approach from the work of \cite{wd01_ion}, so that the average charge of small grains is about 1 for $U>10^3$ and 0 for lower RF intensities. Evolution of big grains is insensitive to the charge as their optical properties are  the same for charged and neutral states. In Fig.~\ref{frac_destr} (left) blue lines show the dependence of the fractional dust loss from a 5~\AA{} size bin on the radiation field intensity. At $U<10^2$~grains of this size are not destroyed at all for 1~Myr. At higher intensities the fraction of destroyed dust, $f_{\rm destr}$, starts to grow, reaching 100\% at $U=10^4$. Aromatisation process is more efficient. The dependencies of fractions of fully aromatised dust grains with radii of 5, 50 and 1000~\AA{} on the radiation field intensity are shown by green, red, and cyan lines, correspondingly. The 50~\AA{}-grains are aromatised independently on the value of $U$ (in the considered range), other grains are fully aromatised at $U>10$.

In Fig.~\ref{frac_destr} (right) the infrared dust emission spectra after 1 Myr of evolution are presented. Spectra were computed using {\tt  DustEm}\footnote{https://www.ias.u-psud.fr/DUSTEM/index.html} tool~\citep{dustem}. The optical properties for the HAC grains with different $E_{\rm g}$ as well as grain size distributions were taken from this work, while heat capacities are assumed to be the same as those for HAC grains in J13 model. The infrared bands are most sensitive to environmental conditions. At $U<10^2$, the band near 3~$\mu$m corresponds to vibrations of aliphatic bonds, at $U\geqslant10^2$ the contribution of aromatic bonds starts to dominate, and in this case the band at 3.3~$\mu$m is more intense than that at 3.4~$\mu$m. At intensities $U \gtrsim 5\cdot 10^3$ the 3~$\mu$m band is barely seen, because the smallest grains responsible for this band are destroyed.

Bands around 6--8~$\mu$m attain features more typical for aromatic grains at lower intensities than bands around 3~$\mu$m. At $U=5$, the spectrum in the 6--8~$\mu$m range corresponds to the mix of aliphatic and aromatic grains but at higher intensities only bands characteristic of aromatic grains remain. More rapid aromatisation of bands in the 6--8~$\mu$m range is related to faster aromatisation of bigger grains, which are responsible for the emission at these wavelengths. Moreover, these aromatic bands do not completely disappear even at the highest considered intensities, because the grains are too big to be destroyed by the UV radiation in the current model.

\begin{figure}
	\includegraphics[width=0.4\textwidth]{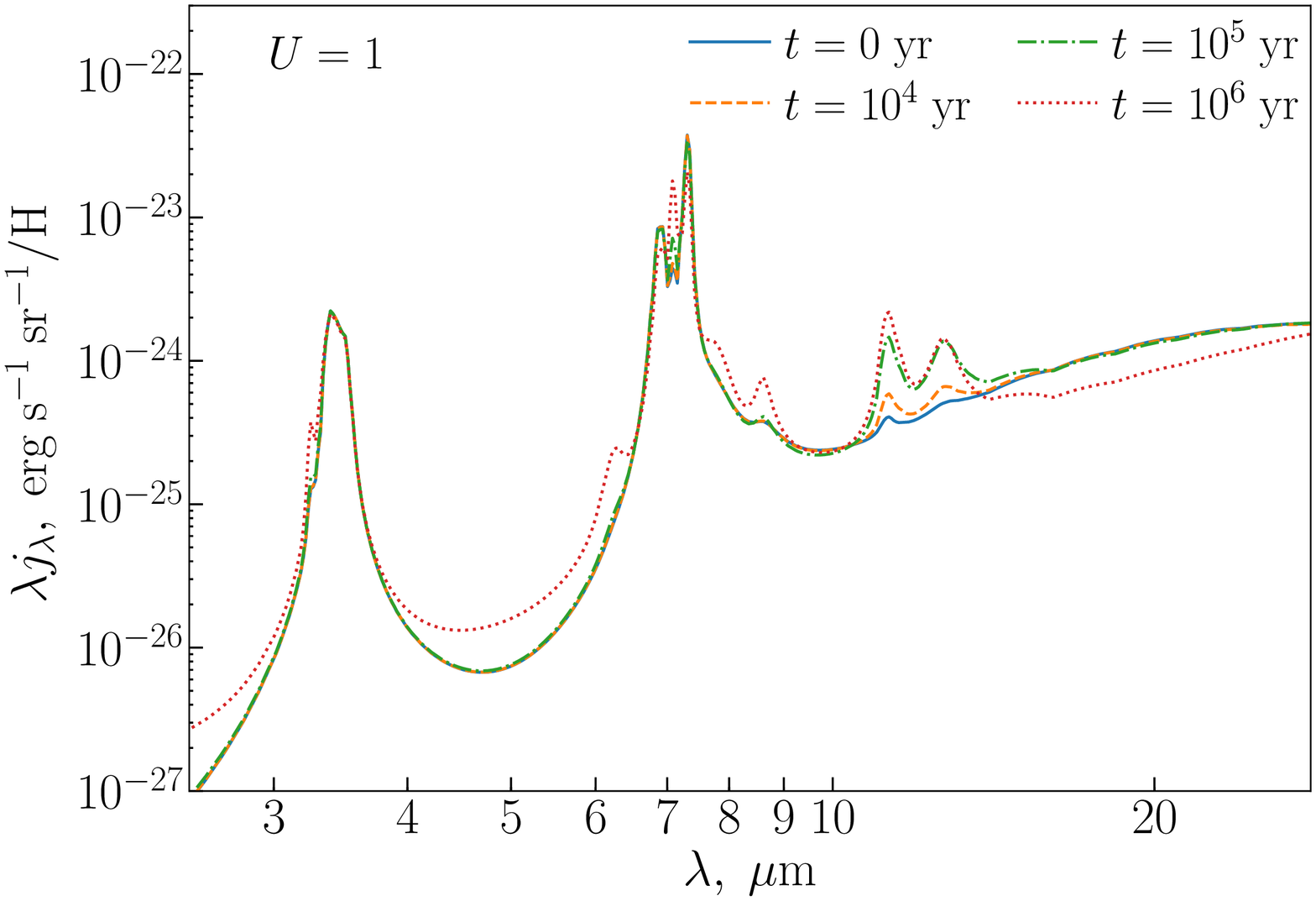}
	\includegraphics[width=0.4\textwidth]{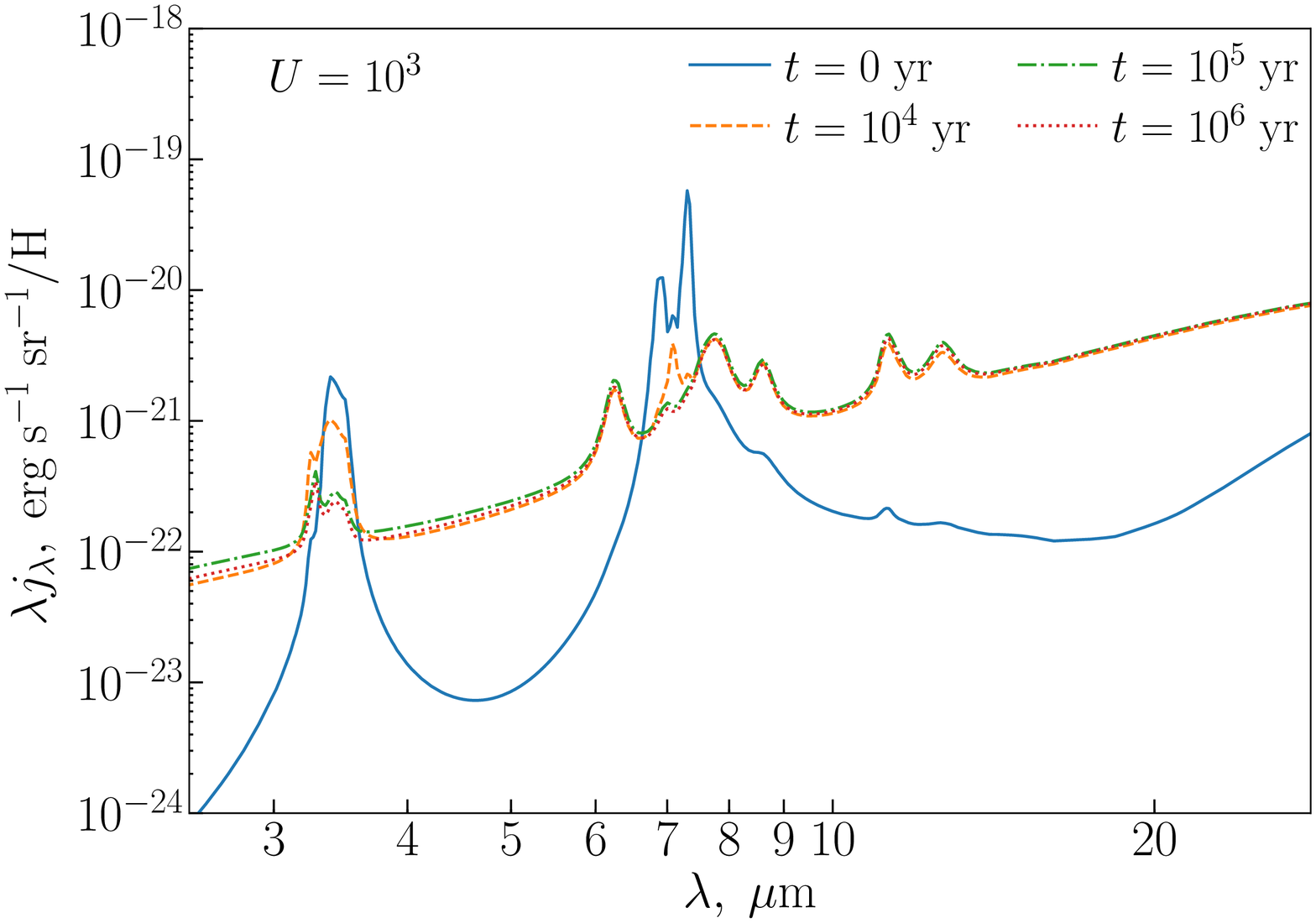}
	\includegraphics[width=0.4\textwidth]{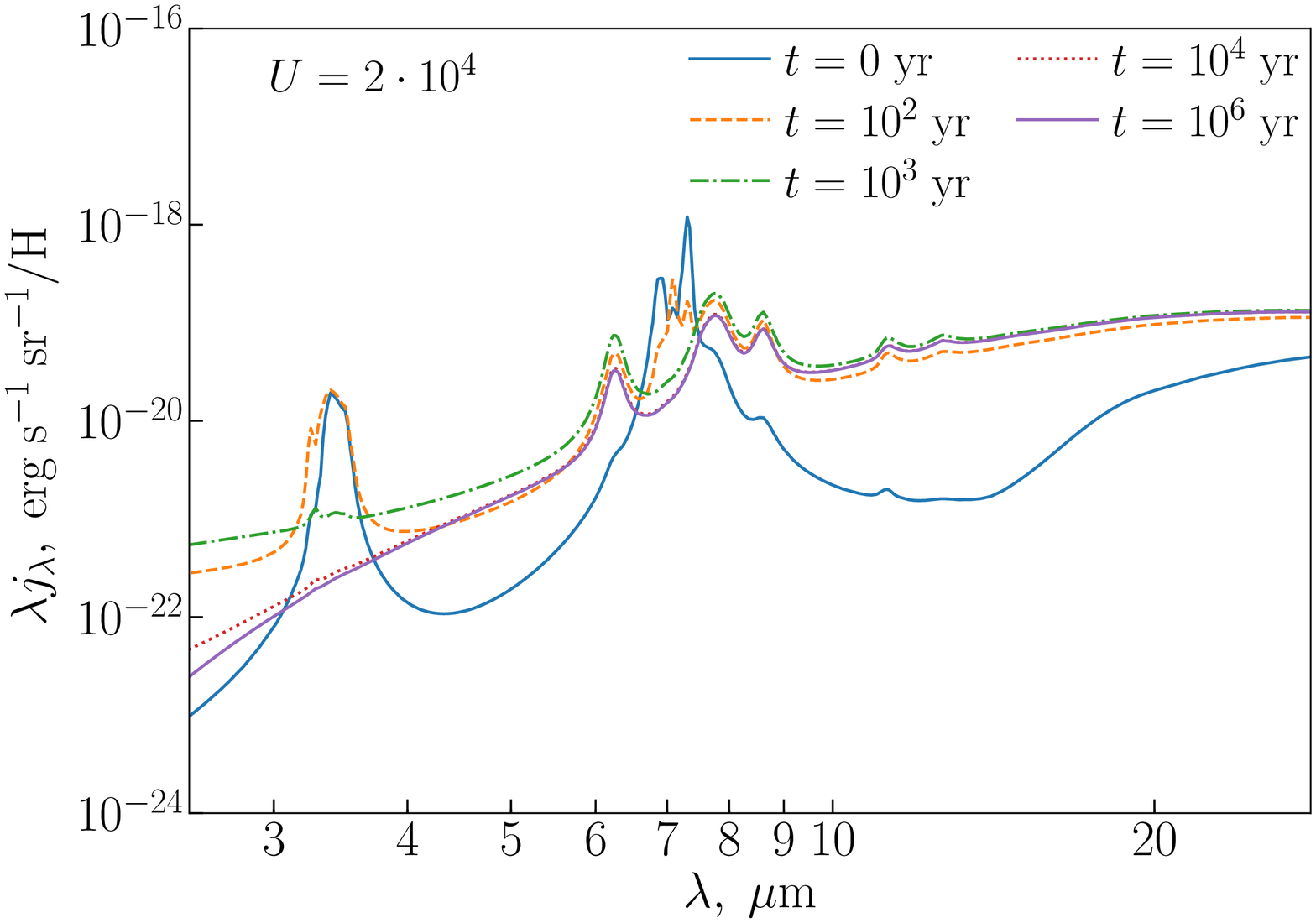}
	\includegraphics[width=0.4\textwidth]{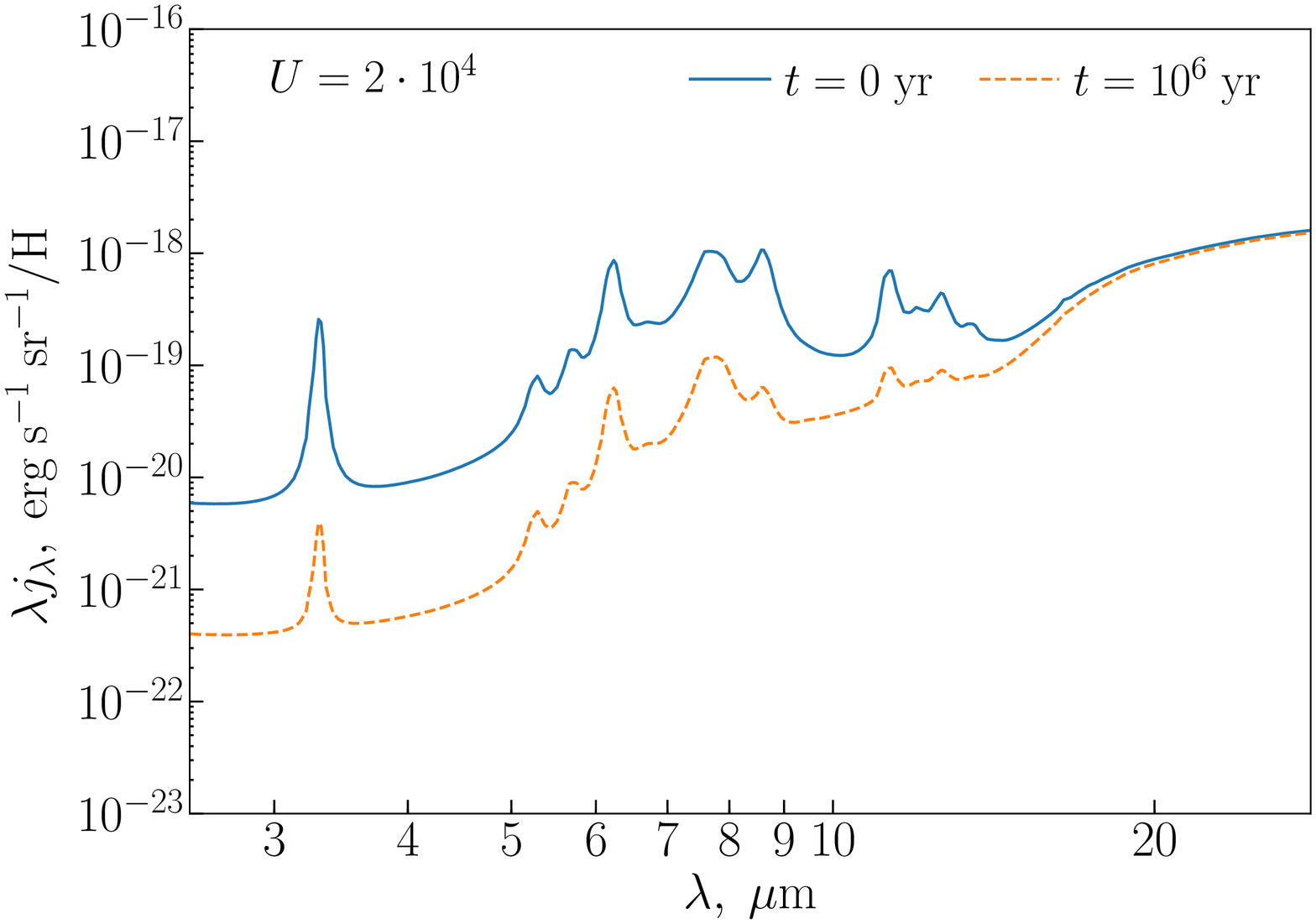}
	\caption{Evolution of infrared spectra at different RF intensities with parameters from J13 and DL07 models. The value of $U$ and the adopted size distribution from top to bottom are $U=1$ and J13, $U=10^3$ and J13,  $U=2\cdot10^4$ and J13, $U=2\cdot10^4$ and WD01. Colours indicate the model time.}
	\label{spectra_pdr}
\end{figure}

We illustrate the evolution of infrared spectra at different RF intensities in Fig.~\ref{spectra_pdr}. At $U=1$, the spectrum near 3~$\mu$m remains almost invariable during the considered time span, while at longer wavelengths (around 10--20~$\mu$m) the spectrum changes due to aromatisation of medium-sized grains. At higher irradiation intensities the spectra at 3~$\mu$m also change as the smallest grains become aromatic, too. At $U=10^3$, the transition from aliphatic-dominant spectrum to aromatic-dominant spectrum occurs between $10^4$ and $10^5$~yr, when the band at 3.3~$\mu$m becomes more intense than the band at 3.4~$\mu$m. At $U=2\cdot10^4$ this transition occurs faster by approximately 2 orders of magnitude, and the bands near 3~$\mu$m disappear during further evolution due to destruction of corresponding grains. We chose the specific value of $U=2\cdot10^4$ as this intensity is representative for the Orion Bar \citep{tielens85, goicochea15}. Bands at both 3.3 and 3.4~$\mu$m are observed in the Bar, although they are not prominent. On the other hand, the bands near 11.2 and 12.6~$\mu$m are strong in the Bar, while, according to our calculations, these bands should not be so intense at this RF value. Thus, our model indicates that properties of amorphous carbons alone are not sufficient to explain emission around 11\,$\mu$m in this region, and some other processes are important in addition to those already included in the model. In fact, at such intense radiation field amorphous carbonaceous grains likely fragment to slices forming PAHs or small hydrocarbons. The inconsistency between our model and the Orion Bar observations can be relaxed if we replace or amend HAC optical properties with the properties of genuine PAHs.

This is further illustrated in Fig.~\ref{spectra_pdr} (bottom), where we show evolving infrared spectra at $U=2\cdot10^4$ for an initial size distribution from the work of \cite{wd01} (WD01) and the optical properties from the work of \cite{DL07} (DL07). This distribution explicitly contains PAHs with their corresponding optical properties that we use to compute both photo-rates and infrared spectra. Indeed, in this case we do see all the PAH infrared bands. Obviously, they also become less intense with time, but still exist after 1~Myr of evolution. This confirms that there should be an additional step in the evolution of HAC particles, when fragments of HAC have optical properties similar to that of planar aromatic macromolecules like PAHs. In the current version of the model we do not consider this step and the corresponding evolution of optical properties.

\subsubsection{Photo-processing in IR bubbles}

Another astrophysical question that can be answered with the {\tt Shiva} model is related to non-uniform spatial dust emission in the so-called IR bubbles~\citep{anderson12}. Most of these objects are believed to be expanding H{\sc ii} regions around massive stars surrounded by gas-dust clouds. Their name stems from a ring-like structure of the dust emission in these objects. While big grains can be relatively easy swept by radiation pressure~\citep{draine11,2015MNRAS.449..440A}, small nano-size grains cannot be easily blown-out from the inner part of the ionized region. Some other factor is needed to explain the lack of small grains there. The radiation field in the star vicinity is very strong, so one may suggest that small grains are efficiently destroyed by UV photons. The {\tt Shiva} model provides an easy way to check this hypothesis and show how IR spectra change during the H{\sc ii} region evolution.

As a reference for IR bubble parameters, we use results of numerical modelling, pre-computed with the MARION code and presented in the works of \cite{pavyar13} and \cite{2015MNRAS.449..440A, 2017MNRAS.469..630A}. To track the changes in the object physical structure, we choose three representative locations within the modelled object. One of this location approximately corresponds to the inner wall of the dense shell that envelops the ionized region. Specifically, it is chosen to be the location where the density of neutral atomic hydrogen reaches its maximum ($n_{\rm H}^{\max}$). The second location corresponds to the density of neutral hydrogen, which is about half of its maximum value ($0.5n_{\rm H}^{\max}$). This is roughly a transition from neutral hydrogen to ionized hydrogen. Finally, the third location is selected inside the H{\sc ii} region, at the point, where density of ionized hydrogen has its maximum ($n_{\rm H+}^{\max}$). As the H{\sc ii} region expands, these locations also move away from the star. At each time moment from 0 to about 300 kyr, we extract physical parameters for these locations, including the radiation field, and use them to simulate dust evolution with the {\tt Shiva} code.

\begin{figure}
	\includegraphics[width=0.45\textwidth]{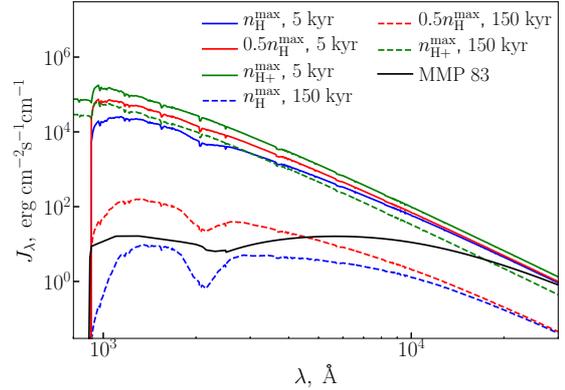}
	\caption{Dependence of radiation field intensity on wavelength for three representative locations considered in the work at 5 and 150~kyr. The MMP~83~\citep{MMP83} radiation field is also presented.}
	\label{field_rcw}
\end{figure}

Radiation field intensities for two time moments, 5 and 150~kyr, at the considered locations are illustrated in Fig.~\ref{field_rcw}. For comparison, the MMP83 radiation field is also shown. At 5~kyr the radiation field intensity at all points is very strong, 4--5 orders higher than MMP83. Also, at the point inside the H{\sc ii} region there are photons with $\lambda<912$~\AA{}, which are absent at other locations. At 150~kyr the radiation field intensity remains high only inside the H{\sc ii} region, while the field at other points becomes relatively weak and comparable to MMP83.

\begin{figure*}
	\includegraphics[width=0.9\textwidth]{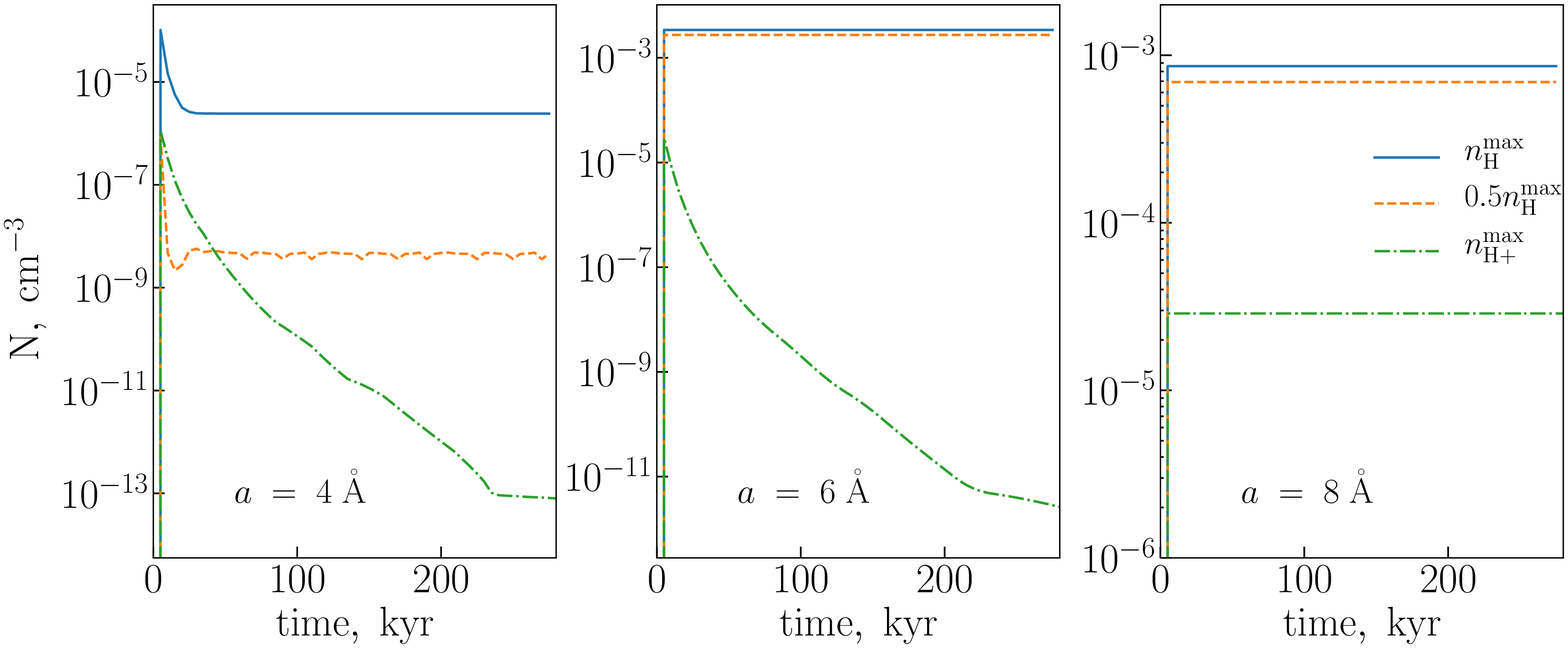}
	\caption{Evolution of number density of dust grains with radii of 4~\AA{} (left panel), 6~\AA{} (middle panel) and 8~\AA{} (right panel). Blue colour corresponds to the location with the maximum density of neutral hydrogen, orange colour corresponds to the location with the density of neutral hydrogen $0.5n_{\rm H}^{\rm \max}$, and green colour corresponds to the location with the maximum density of ionized hydrogen.}
	\label{evol_rcw}
\end{figure*}

We adopt the WD01 initial size distribution and assume that initially dust grains are fully hydrogenated HACs. In Fig.~\ref{evol_rcw} the evolution of number density of dust grains in the size bins corresponding to radii of 4, 6, and 8~\AA{} at three selected locations is presented. Note that we only show the density of dehydrogenated grains. While initially their density is zero (because all grains are hydrogenated), but all grains become fully dehydrogenated during first 5~kyr, and afterwards they can lose only carbon atoms. The smallest grains experience significant destruction at all the considered locations, but only in the inner part of the H{\sc ii} region destruction of 4~\AA{}-grains occurs over the entire calculation time, 300~kyr. Outside the inner part of the ionized region destruction practically stops after $\sim20$~kyr. Due to the region expansion, the radiation field becomes too weak for photo-destruction there.

Grains with a radius of 6~\AA{} are destroyed in the H{\sc ii} region, but stay mostly intact in more remote locations (with half-maximum and maximum neutral hydrogen density). The 8~\AA{} grains are not destroyed anywhere. This means that in spite of very strong radiation field around massive stars only the smallest grains can be photo-destroyed within H{\sc ii} regions, at least, according to the evolutionary model implemented in {\tt Shiva}. We use the RRK approach to estimate the dissociation rate, and this rate strongly depends on the number of carbon atoms in a grain. This causes large decrease in the dissociation rate, when the size of a grain varies only a little.

\begin{figure*}
	\includegraphics[width=0.9\textwidth]{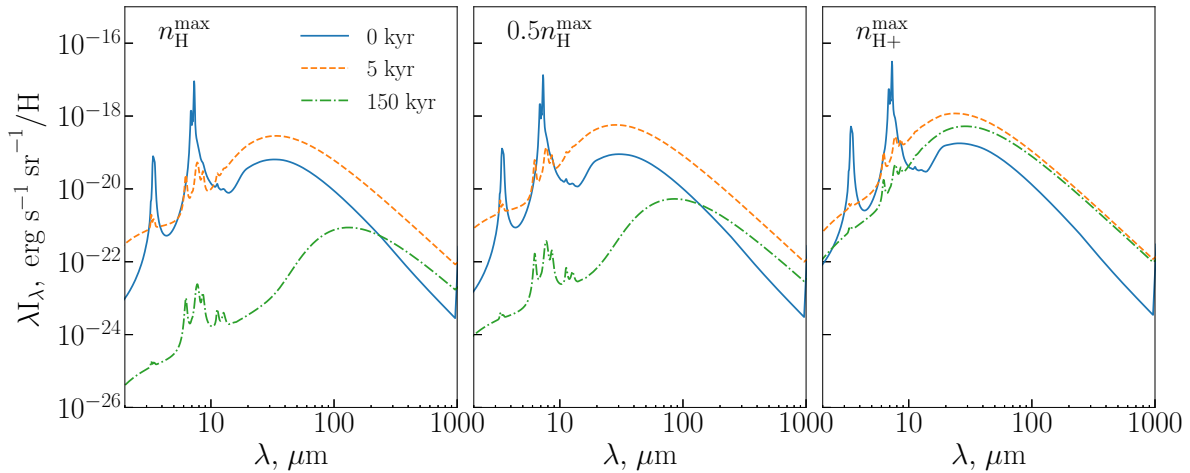}
	\caption{Infrared spectra for various locations within the expanding H{\sc ii} region: left panel --- the point of the maximum neutral hydrogen density, middle panel -- the point where neutral hydrogen density is a half of its maximum value, right panel -- the point of the maximum ionized hydrogen density. Spectra for three time moments are shown in each panel: blue curves --- 0~kyr, orange curves --- 5~kyr, green curves --- 150~kyr.}
	\label{spec_rcw}
\end{figure*}

Let us now consider how this evolution of small grains influences infrared spectra for the whole grain ensemble, shown in Fig.~\ref{spec_rcw} for the three locations described above. As initially all dust grains are hydrogenated, the spectra at $t=0$ shown with blue lines in all three panels demonstrate IR features corresponding to fully hydrogenated amorphous carbon grains. After only 5 kyr of the evolution the spectra change: the features of aliphatic bonds disappear, while IR features corresponding to aromatic bonds emerge in the spectrum. First they are quite strong, but later gradually weaken, especially at 3~$\mu$m, because this band arises from the smallest grains, which are destroyed over the entire region. Features at 6--8~$\mu$m only slightly stand out above the continuum spectrum of hot large grains in the right panel, i.e. in the ionized hydrogen region, as 6~\AA{} grains are mostly destroyed there. At 150~kyr the spectra only change due to the radiation field change, but the number density of dust grains at the selected locations does not vary significantly since then.

Summarising, in a typical IR bubble small grains with radius of about 4~\AA{} cannot survive. Larger grains with radius 6~\AA{} can survive only close to the dense shell, where neutral hydrogen dominates and there are no photons with energy higher than 13.6~eV. Grains with radius of 8~\AA{} can survive even inside the bubble, where the ionized hydrogen dominates. As no emission at 8~$\mu$m is observed in the central parts of real bubbles, there should be some other mechanism, responsible for the absence of 8~\AA{} grains in the centres of infrared bubbles.

\subsection{Evolution of grains in the medium with high gas/dust velocities and temperature}\label{subsect: SNR}

To demonstrate the evolution of dust in the medium with high velocities and temperatures we have performed calculations on a grid of velocities $v_{\rm gas}$ in the range of $[30...200]$~km~s$^{-1}$ and temperatures $T_{\rm gas}$ in the range of $[10^{3}...10^6]$~K with a fixed value for a product $n_{\rm H}T_{\rm gas}=10^4$~cm$^{-3}$~K, taking into account that temperature and density can be coupled and anti-correlated. Abundances of helium and carbon are assumed to be $n_{\rm He} = 0.1 n_{\rm H}$ and $n_{\rm C} = 10^{-4} n_{\rm H}$, respectively. We deliberately assume zero RF in these computations to highlight the effects of sputtering and shattering. The gas-dust relative velocity ($v_{\rm gd}$) is taken to be $3/4$ of the gas velocity, which is appropriate for adiabatic shock waves. We estimate the grain-grain collision velocity between bins $i_1$ and $i_2$ as
\begin{equation}
v^{\rm col}_{i_1i_2} = (\sigma_{{\rm d},i_1}^{2}+\sigma_{{\rm d},i_2}^{2})^{1/2},
\end{equation}
where $\sigma_{\rm d}$ is grain velocity standard deviation, which is equal to the gas-dust relative velocity. It is also assumed that dust differential drift is negligible.

\begin{figure*}
	\includegraphics[width=0.9\textwidth]{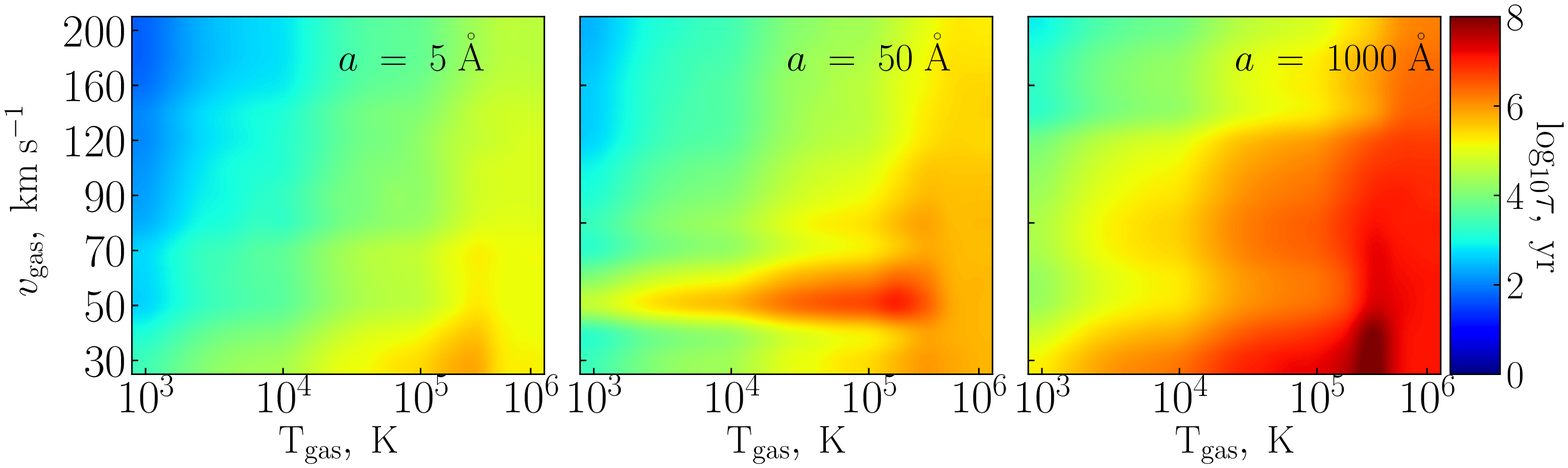}
	\includegraphics[width=0.9\textwidth]{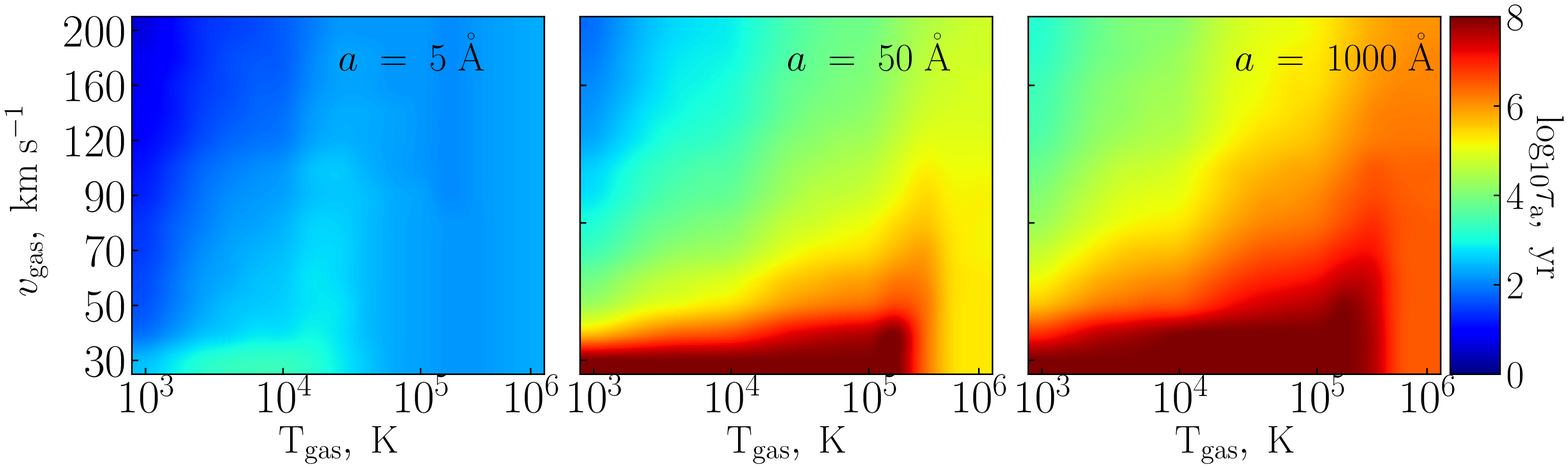}
	\includegraphics[width=0.9\textwidth]{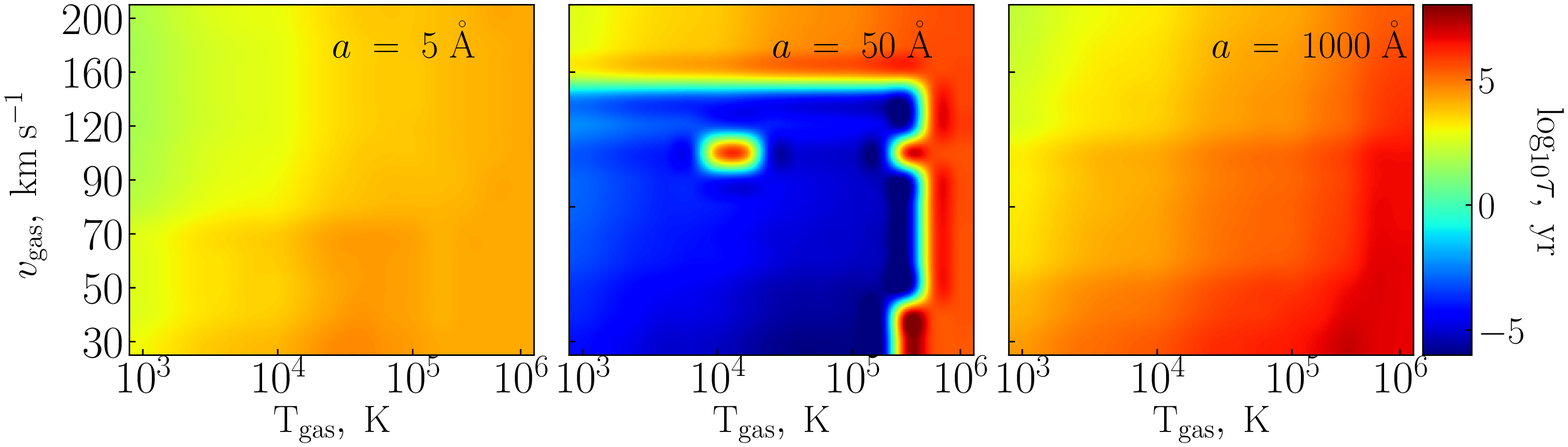}
	\caption{Top row: time-scales $\tau$ of grain destruction for grains with initial radii of 5, 50 and 1000 \AA{} as functions of the gas velocity and temperature assuming the initial size distribution from the work of \citet{jones13}. Middle row: time-scales $\tau_{\rm a}$ of aromatisation for grains of the same initial sizes as in the top row. Bottom row: the same as in the top row, but for the WD01 initial size distribution. Note that in the bottom row notation `$-6$' means negative time-scale $\tau=-10^6$~yr, i.e. dust production.}
	\label{timescales}
\end{figure*}

In Fig.~\ref{timescales} (top row) we demonstrate maps of grain destruction time-scales $\tau$ for various grain sizes. We stop the calculations, when the ratio $\sum\limits_{j}N_{ij}(0)$/$\sum\limits_{j}N_{ij}(t)$ is around $e$. In the simulations an exact $e$ value is never reached, and this may introduce some minor uncertainties in determining the time-scales, but they do not alter the overall picture. In the top and middle rows of this diagram the results are presented for amorphous carbon grains with the J13 size distribution, which are assumed to be fully hydrogenated initially. The grains with radii of 5, 50 and 1000~\AA{} are chosen as typical representatives of the smallest nano-size grains, very small grains, and big grains, correspondingly.

The time-scale for each grain type is a result of the balance of the considered processes. In Fig.~\ref{timescales} (top row) for 5\AA{}-grain the destruction time-scale gradually increases from high velocities and low temperatures (high density) to low velocities and high temperatures (low density) reflecting that the density plays a key role in determining the destruction efficiency. The same trend is seen for grains with radii of about 50~\AA{} and 1000 \AA{}. This implies that, despite the sputtering process being very efficient at high temperature, grains have large time-scales due to low gas density. The decreasing time-scale can only be noted for highest values of considered temperatures. Presence of various heterogeneities on the maps clearly shows how subtle the balance of evolutionary processes is and how strongly it depends on the adopted conditions.

Smoother changes are seen for aromatisation time-scales in Fig.~\ref{timescales} (middle row) as the only process that influences this time-scale is sputtering. The most efficient aromatisation occurs at the highest velocities and again it slows down with the decreasing gas density.

As the outcome of the evolution strongly depends on the abundance of small grains, it is obviously sensitive to the adopted initial size distribution. In the bottom row of Fig.~\ref{timescales} we show the same map of time-scales as in the top row, but for graphites and the WD01 initial size distribution.

In the case of the WD01 initial size distribution, the number density in a specific mass bin can both increase and decrease during the evolution, which we denote with the sign of the corresponding time-scales. Negative values indicate increasing number density, and this may indeed happen with smaller grains at the considered conditions. Specifically, the replenishment of 50~\AA{}-grains due to shattering of bigger grains dominates over their destruction via sputtering, when $v_{\rm gas}\lesssim150$~km~s$^{-1}$ and $T_{\rm gas}\lesssim5\cdot10^5$~K. Under different conditions the destruction dominates, and the time-scales are positive. The smallest and the largest grains have only positive time-scales. Destruction time-scales for them always decrease with velocity and increase with temperature as in the adopted setup growing temperature implies decreasing density.

While grain destruction by energetic particles has already been considered in the literature, the novel feature of our model is that we also consider simultaneous changes in the grain structure. We demonstrate the efficiency of aromatisation by energetic ions and electrons, using the results of supernova remnant (SNR) shock simulations performed in the work of \cite{nozawa06}. The sputtering process was treated in their calculations, so we simulate only aromatisation process adopting the evolution of grain sizes from their work. Calculations by \cite{nozawa06} are intended to reproduce conditions in the early Universe, but this is not a critical issue for our study as we only use their results as a reference for a sample computation.

Initial shock velocity is about 3000~km~s$^{-1}$, but the starting time of dust evolution calculation is 2000~yr in their model. By this time the shock velocity falls to $\sim1000$~km~s$^{-1}$, temperature of the shock front is $\sim6\cdot10^7$~K, and gas number density is $\sim3$~cm$^{-3}$. As dust grains of different radii move differently, the conditions for them at each time also differ. Distance, over which a grain moves inside the shock, varies depending on the time-scale of its destruction. The smallest grains are destroyed rapidly, while large grains are not completely destroyed and gradually turn around in the direction of shock wave motion. So the conditions (relative gas-dust velocity, density, temperature) cover different ranges for different grain sizes. A dust grain with a radius of $a=8.6\cdot10^{-7}$~cm is exposed to the following conditions: $v_{\rm gd}$ in $[1000\ldots15]$~km~s$^{-1}$, $T_{\rm gas}$ in $[6\cdot10^7\ldots4\cdot10^7]$~K and $n_{\rm H}$ in $[3-1]$~cm$^{-3}$, where the ranges are defined by the lifetime of the grains. Grains of larger radii ($a>1.7\cdot10^{-6}$~cm) survive in the shock, so ranges of $v_{\rm gd}$, $T_{\rm gas}$ and $n_{\rm H}$ for them are wider:  $[1000\ldots0]$~km~s$^{-1}$, $[6\cdot10^7\ldots1\cdot10^7]$~K, $[3-0.1]$~cm$^{-3}$, correspondingly.

\begin{figure*}
	\includegraphics[width=0.45\textwidth]{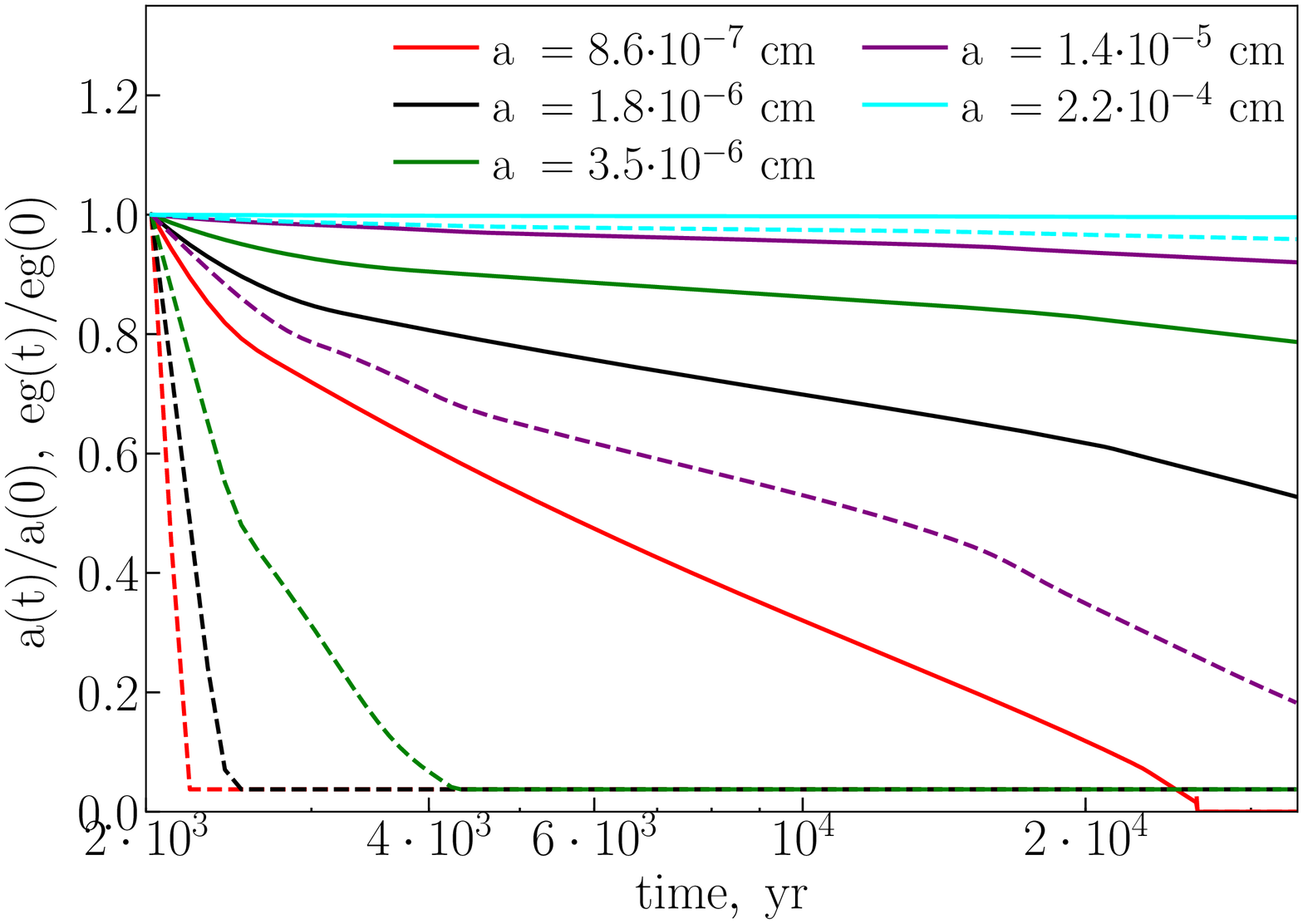}
	\includegraphics[width=0.45\textwidth]{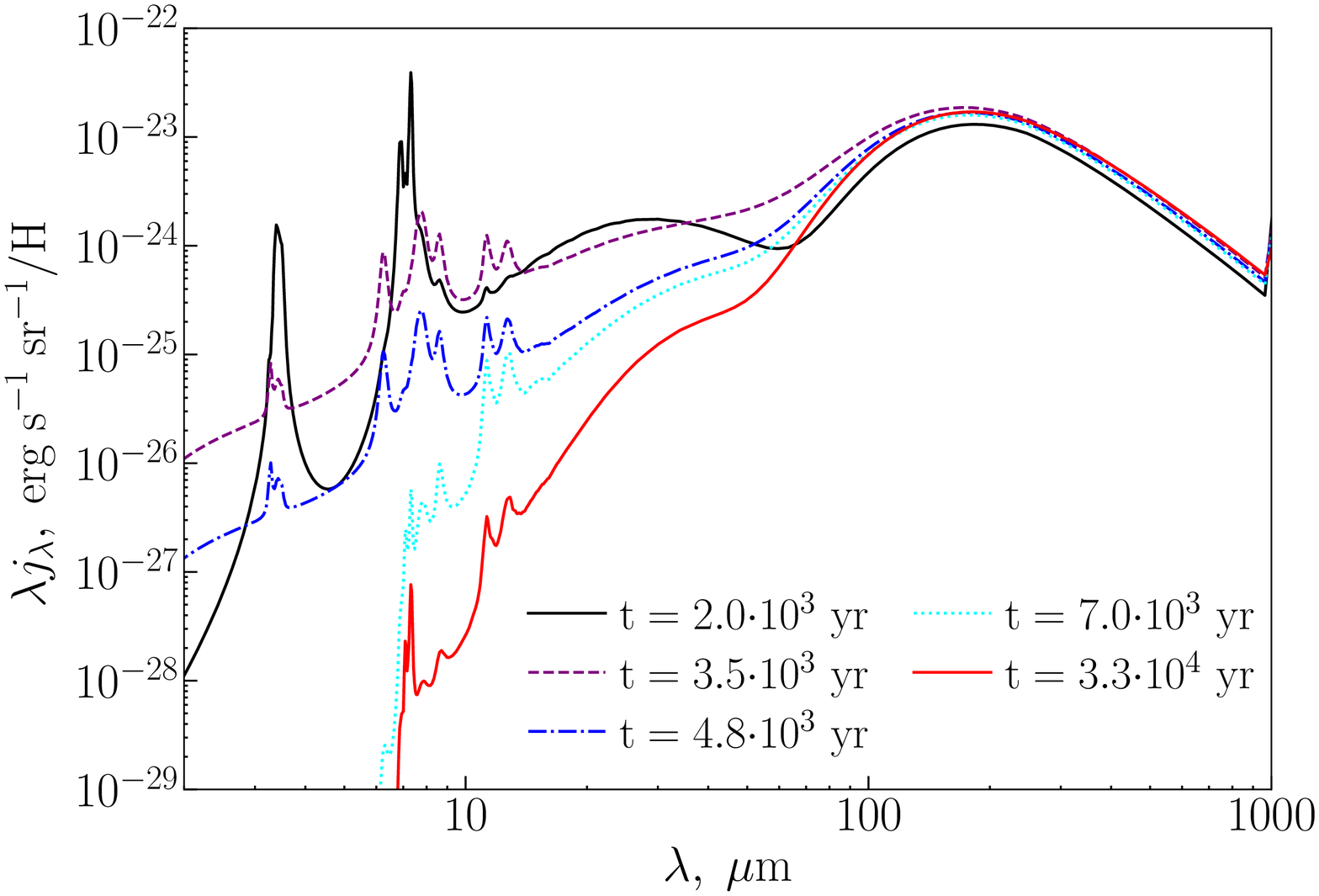}
	\caption{Left: evolution of grain radius \citep[solid line, taken from][]{nozawa06} and band gap energy (dashed line, this work) for different initial grain sizes. Right: the infrared spectra of dust in the SNR shock wave region at different time steps. The starting time of the calculation is 2$\cdot10^3$~yr.}
	\label{SNR_arom}
\end{figure*}

In Fig.~\ref{SNR_arom} (left) we show how various grains change their radii (solid lines) and band gap energies (dashed lines) as they traverse the shock. A time-scale of aromatisation is always smaller than a grain destruction time-scale. Grains with radii of about $8.6\cdot10^{-7}$~cm and smaller disappear after the supernova shock wave passage. Thus, the aromatisation level of these grains is irrelevant. Larger grains survive in the shock wave though their radii are reduced. Grains smaller than $\approx 1.4 \cdot 10^{-5}$~cm are fully aromatised by ions and electrons, while larger grains remain aromatised only partially.

In Fig.~\ref{SNR_arom} (right) we present the infrared spectra of the modelled SNR shock region, integrated spatially to include grains of all sizes, which have different velocities and positions in the considered model. Initially the spectrum is that of fully hydrogenated HACs (black line). By the time of $3.5\cdot10^3$~yr small grains are completely aromatised, and the mid-IR spectrum has the shape typical for aromatic grains. Later, at $4.8\cdot10^3$~yr, number density of small grains becomes smaller, so the emission intensity in the wavelength range of 2--50~$\mu$m decreases almost by the order of magnitude. At even later times small dust grains are destroyed, and the emission in the mid-IR range is nearly absent. Only large grains contribute to the spectrum at longer wavelengths. Such computations can be useful for interpreting real observations of the IR spectrum evolution across the SNR shock, like those reported by \citet{2006ApJ...653..267T} and \citet{2010ApJ...725..585A}. Specifically, comparing observed spectra and {\tt Shiva} predictions, we can infer the time-scale of the supernova remnant expansion similarly to what has been done by \citet{2012ApJ...754..132T}. Also, it may turn out that the fading brightness of the {\em Spitzer} 3.6 $\mu$m band observed in the remnant of SN1987A \citep{2016AJ....151...62A} is at least partially caused by grain restructuring.

The {\tt Shiva} tool is well suited to provide quick preliminary answers to these questions, which will be useful for interpreting SOFIA and forthcoming JWST spectral observations. Other possible scenarios of {\tt Shiva} usage may include studies of grain evolution in the intergalactic medium \citep{intergalactic}, establishing the origin for various correlations in gas and dust properties in star-forming regions \citep{sm01,sm02}, etc.

\section{Summary}\label{sect: summary}

Dust evolution in the ISM is a hot topic in the modern astrophysics, and it is important even in those studies, which are not directly concerned with the dust itself. We present a physical model {\tt Shiva} for simulating the dust evolution in {\rm warm and hot ISM phases}. The model is implemented as a numerical tool that can be used for estimating the evolutionary time-scales for various grains at different conditions in the ISM and for studying dust size evolution. The main destructive processes included in the presented version of the model are photo-processing, shattering, and sputtering. The input model parameters are the radiation flux, hydrogen density (and also densities of helium and carbon as well as fractions of ionized H, He, C), gas velocity and temperature, dust velocity and its dispersion. It is also possible to change the size distribution and type of dust: silicates, HAC grains or graphites+PAHs. The presented numerical tool is publicly available.

We have performed test calculations and presented the dependence of characteristic time-scales on radiation field, gas velocity and temperature. For the last two parameters we constructed the maps of destruction and aromatisation time-scales. These calculations show that the dust evolution is a complicated process and that photo-destruction, sputtering and shattering must be considered simultaneously. We also demonstrated evolutionary changes of the infrared dust emission spectra  due to destruction and aromatisation processes in the media with enhanced radiation field and with high gas velocity. Beside the conditions in a PDR and a supernova shock, we also applied our model to the conditions in an IR bubble around an expanding H{\sc ii} region and show how the model can be used to estimate time-scales of dust evolution in these objects.

We have considered the influence of dust charge on the evolutionary processes and concluded that the charge influences the photo-destruction process. Changes in ionization potential and cooling rate for charged dust grains leads to their faster destruction in comparison to neutral grains. 

The present model does not account for hydrogenation of HACs and their subsequent aliphatisation, volatile accretion and coagulation processes that dominate in the cold neutral medium, so it cannot be used for modelling molecular clouds or protoplanetary discs, where these processes are crucial. 

The authors are very grateful to Anthony Jones for interest to this work, fruitful discussions and comments. We also thank Takaya Nozawa for providing the shock profiles and Sergey Khoperskov for stimulating discussions. The CVODE package (SUNDIALS software) was used in the work and we express our gratitude to the SUNDIAL team for providing this tool. The work was supported by the RFBR grants 17-02-00521 and 18-52-52006. 
\bibliographystyle{mnras} 
\bibliography{main}

\appendix

\section{Sputtering rate coefficients}
\label{app: sput_coef}

A rate coefficient of aromatisation due to inelastic collision with thermal and non-thermal ions (H, He, C) and electrons for grains with less than 1000~atoms is calculated as follows:
\begin{equation}
\begin{split}
\varepsilon^{(2)}_{ij} &= \frac{\delta}{N_{\rm atoms}} \left[R_{\rm s}^{\rm H}({\rm H}) + R_{\rm s}^{\rm He}({\rm H})+ R_{\rm s}^{\rm C}({\rm H}) \right. \\ 
&\left. + R_{\rm s,T}^{\rm H}({\rm H}) + R_{\rm s,T}^{\rm He}({\rm H})+ R_{\rm s,T}^{\rm C}({\rm H}) +R_{\rm s,T}^{\rm elec}({\rm H})\right].
\end{split}
\end{equation}
A rate coefficient of sputtering of carbon atoms due to inelastic collision with thermal and non-thermal ions (H, He, C) and electrons for grains with less than 1000~atoms is calculated as follows:
\begin{equation}
\begin{split}
\mu^{(2)}_{ij} &= \frac{\mu_{\rm d}}{N_{\rm A}}\left[R_{\rm s}^{\rm H}({\rm C}) + R_{\rm s}^{\rm He}({\rm C})+ R_{\rm s}^{\rm C}({\rm C})\right.\\ 
&\left. + R_{\rm s,T}^{\rm H}({\rm C}) + R_{\rm s,T}^{\rm He}({\rm C})+ R_{\rm s,T}^{\rm C}({\rm C}) +R_{\rm s,T}^{\rm elec}({\rm C})\right].
\end{split}
\end{equation}

A rate coefficient of aromatisation due to elastic collision with thermal and non-thermal ions (H, He, C) for grains with less than 1000~atoms is calculated as follows:
\begin{equation}
\begin{split}
\varepsilon^{(3)}_{ij} &= \frac{\delta}{N_{\rm atoms}}\left[R_{\rm n}^{\rm H}({\rm H}) + R_{\rm n}^{\rm He}({\rm H})+ R_{\rm n}^{\rm C}({\rm H}) \right. \\ 
&\left. + R_{\rm n, T}^{\rm H}({\rm H}) + R_{\rm n, T}^{\rm He}({\rm H})+ R_{\rm n,T}^{\rm C}({\rm H})\right].
\end{split}
\end{equation}
A rate coefficient of sputtering of carbon atoms due to elastic collision with thermal and non-thermal ions (H, He, C) for grains with less than 1000~atoms is calculated as follows:
\begin{equation}
\begin{split}
\mu^{(3)}_{ij} &= \frac{\mu_{\rm d}}{N_{\rm A}} \left[R_{\rm n}^{\rm H}({\rm C}) + R_{\rm n}^{\rm He}({\rm C})+ R_{\rm n}^{\rm C}({\rm C}) \right. \\ 
&\left. + R_{\rm n,T}^{\rm H}({\rm C}) + R_{\rm n,T}^{\rm He}({\rm C})+ R_{\rm n,T}^{\rm C}({\rm C})\right].\\
\end{split}
\end{equation}

A rate coefficient of aromatisation due to elastic collision with thermal and non-thermal ions (H, He, C) for grains with more than 1000~atoms is calculated as follows:
\begin{equation}
\begin{split}
\varepsilon^{(4)}_{ij} &= \frac{\delta}{N_{\rm atoms}} \left[R_{\rm n}^{\rm H}({\rm H}) + R_{\rm n}^{\rm He}({\rm H})+ R_{\rm n}^{\rm C}({\rm H})\right. \\ 
&\left. + R_{\rm n, T}^{\rm H}({\rm H}) + R_{\rm n, T}^{\rm He}({\rm H})+ R_{\rm n,T}^{\rm C}({\rm H})\right].
\end{split}
\end{equation}
A rate coefficient of sputtering of carbon atoms due to elastic collision with thermal and non-thermal ions (H, He, C) for grains with more than 1000~atoms is calculated as follows:
\begin{equation}
\begin{split}
\mu^{(4)}_{ij} &= \frac{\mu_{\rm d}}{N_{\rm A}} \left[R_{\rm big}^{\rm H}({\rm C}) + R_{\rm big}^{\rm He}({\rm C}) + R_{\rm big,T}^{\rm H}({\rm C}) + R_{\rm big,T}^{\rm He}({\rm C})\right].\\
\end{split}
\end{equation}

Expressions for ejection rates of C and H atoms were given in Paper~I.

\section{Photo-destruction rate coefficients} 
\label{app: photo_coef}
A rate of change of band gap energy $\varepsilon^{(1)}_{ij}$ due to aromatisation by photons can be expressed as: 
\begin{equation}
\varepsilon^{(1)}_{ij} = \delta Y_{\rm diss}^{\rm CH} \sigma^{\rm CH}_{\rm loss} \int\limits_{10{\rm eV}}^{13.6{\rm eV}}
Q_{\rm abs}(a_i,\eg^{j}, Z_i, E) \frac{F(E)}{E} dE,
\label{eq: arom}
\end{equation}
where $\delta$ is a conversion coefficient between hydrogen atom fraction $X_{\rm H}$ and band gap energy, $E_{\rm g}=\delta X_{\rm H}$, $Y_{\rm diss}^{\rm CH}$ is the dissociation probability, $\sigma^{\rm CH}_{\rm loss}$ is the C--H dissociation cross section. We adopt $\delta=4.3$~eV~\citep{tamorwu90}. Following \citet{jones14}, we adopt $Y_{\rm diss}^{\rm CH} = 1$ for grains with $a<20$~\AA{} and $Y_{\rm diss}^{\rm CH} = 20$\AA{}$/a$ for larger grains, and $\sigma^{\rm CH}_{\rm loss} = 10^{-19}$~cm$^2$.

In the model of \cite{allain96_1}, which is used to calculate the second step of destruction, when the grain is dehydrogenated, a hydrogen atom, a hydrogen molecule, or an acetylene molecule are detached from a grain with the rates of $k_{\rm H}$, $k_{\rm H_2}$, $k_{\rm C_{2}H_{2}}$,~s$^{-1}$, correspondingly, which are calculated using the RRK (Rice-Ramsperger-Kassel) theory \citep{RRK} based on the experimental data \citep{jochims94}. The sum of these rates is the total rate dissociation coefficient, $k_{\rm diss} = k_{\rm H}+k_{\rm H_2}+k_{\rm C_{2}H_{2}}$. The dissociation probability of C--C bond by a single photon (sp) can be expressed as:
\begin{equation}
Y_{\rm diss}^{\rm CC, sp}(a_i,E, Z_i) = \frac{2 k_{\rm C_{2}H_{2}}[1-Y_{\rm ion}(a_i,E,Z_i)]}{k_{\rm diss}(a_i,E)+k_{\rm IR}(a_i,E,Z_i)},
\end{equation}  
where $k_{\rm IR}$ is a rate coefficient of IR photon emission (see more details in \citetalias{murga16a}), and factor 2 means that two C--C bonds are destroyed after acetylene detachment. We adopt $Y_{\rm ion}$ from the work of \citet{wd01_ion} developed for PAHs and graphite-like grains supposing that ionization yields for dehydrogenated HACs are similar to those for PAHs. While there is no calculated dependence of $Y_{\rm ion}$ for HAC on their size, energy and charge, it was experimentally found that the work function of HAC films is about 4--5~eV depending on the sp$^{2}/{\rm sp}^{3}$ ratio \citep{illie00}. The analogous value for PAHs is 4.4~eV, which is somewhere between the values for the HAC material. In addition, this value almost does not depend on the film thickness if the HAC material includes sp$^{2}$-nanosized clusters \citep{shakerzadeh12}. So, these studies give grounds to believe that our assumption on $Y_{\rm ion}$ of HACs is reasonable.

\bsp	
\label{lastpage}
\end{document}